## *The polarizability model for ferroelectricity in perovskite oxides*

Annette Bussmann-Holder
Max-Planck-Institute for Solid State Research
Heisenbergstr. 1
D-70569 Stuttgart, Germany

This article reviews the polarizability model and its applications to ferroelectric perovskite oxides. The motivation for the introduction of the model is discussed and nonlinear oxygen ion polarizability effects and their lattice dynamical implementation outlined. While a large part of this work is dedicated to results obtained within the self-consistent-phonon approximation (SPA), also nonlinear solutions of the model are handled which are of interest to the physics of relaxor ferroelectrics, domain wall motions, incommensurate phase transitions. The main emphasis is to compare the results of the model with experimental data and to predict novel phenomena.

### 1. Introduction

Ferroelectricity in perovskite oxides has been discovered in 1945, 1946 in $BaTiO_3$ by Wul and Goldman [1, 2]. This discovery had important consequences since it was the first ferroelectric compound without hydrogen bonds. Further, it was the first ferroelectric with more than one ferroelectric phase. It turned out to be chemically and mechanically very stable with ferroelectricity being realized at room temperature. Compared to hydrogen bonded ferroelectrics its structure is rather simple (cubic at high temperature) constituted of three elements only. The preparation of $BaTiO_3$ is easy and also in ceramic form suited for applications. The main building block is the oxygen octahedra surrounding the central transition metal B and the oxygen ions being located face centered in the cubic A sublattice of $ABO_3$. It was soon realized that $BaTiO_3$ can be widely used in applications which initiated an ongoing search for other perovskite oxides. It was the group around B. T. Matthias who discovered most of the nowadays known ferroelectrics. In 1949 $KNbO_3$ and $KTaO_3$ [3] were synthesized followed by $LiNbO_3$ and $LiTaO_3$ [4]. $PbTiO_3$ was found by Shirane et al. in 1950 [5] and is recognized as one of the most important technological materials.

With the discovery of perovskites the theory of the microscopic origin of ferroelectricity had to be revised fundamentally since until 1945 it was believed that hydrogen bonds play a crucial role for this property. Slater suggested in 1950 [6] that ferroelectricity in $ABO_3$ is caused by a cancellation of long range repulsive dipolar forces by short range local attractive forces. These ideas are the basic ingredients for a displacive phase transition as opposed to order / disorder transitions which at that time were believed to be realized in hydrogen bonded compounds. The Slater model was based on the idea that the titanium ion was rattling in its rigid ion cage thus causing an ionic instability. This ansatz caused problems since the role of the rattling ion remained ambiguous. It took another 10 years that Cochran [7] and Anderson [8] suggested that a lattice mode involving all ions could soften (soft mode concept) and lead to the displacive instability. This issue will be addressed in deeper detail later on in this chapter.

On a macroscopic level a more rapid success was achieved where details about ionic displacements, order / disorder or displacive dynamics are irrelevant, and only thermodynamic aspects are considered. Already in 1940 Müller proposed to use a free energy functional [9] which is expanded in powers of polarization and strain thus being able to relate measured quantities with expansion coefficients. This technique was completed by Ginzburg [10] and Devonshire [11] who showed that the paraelectric as well as the ferroelectric phase can be well described within this approach.

From a microscopic point of view a more detailed understanding was gained between 1960 and 1970 when perovskites were investigated by inelastic neutron scattering experiments where clear evidences for the soft mode behavior were obtained emphasizing the role of lattice dynamics for the occurrence of ferroelectricity.

## 2. Motivation

The starting concept is quite general, namely constructing a Hamiltonian where all degrees of freedom are involved: ionic, electronic and electron lattice interactions. Since ferroelectrics are insulators with a rather large band gap, the complexity has been reduced through the adiabatic principle where the electronic and ionic degrees of freedom can be decoupled from each other and the electrons are tightly bound to the ionic cores to be only a function of the ionic coordinates. This allows to effectively include the electronic energy into the lattice potential energy at a mean field level. The resulting Hamiltonian is the sum of the kinetic energy, the harmonic potential energy and a fourth order term which guarantees the overall lattice stability [12]:

$$H_l = \tfrac{1}{2}\sum_{j,j'} P_j^2(l) + \tfrac{1}{2}\Omega_0^2 \sum_{j,l} Q_j^2(l) + \frac{\Gamma_1}{2}\sum_j Q_j^4(l) + \frac{\Gamma_2}{4}\sum_{j \neq j'} Q_j^2(l)Q_{j'}^2(l') \qquad (1)$$

where $P, Q$ are canonically conjugate variables, $Q(l)$ is the linear combination of the displacements of the individual ions in the unit cell $l$ participating in the soft mode, $\Omega_0$ is the harmonic lattice mode frequency and terms proportional to $\Gamma_1, \Gamma_2$ are quartic interactions. Interactions between cells are given by:

$$H_{\text{int}} = \tfrac{1}{2}\sum_{ll',jj'} v_{jj'}(ll') Q_j(l) Q_{j'}(l') \qquad (2)$$

where $v_{jj'}$ is the harmonic intercell interaction. In the long wave length limit $v_{jj'}$ is non-analytic. In order to avoid this behavior a constant value for $q=0$ is generally adopted, which is taken as a free parameter. Typically the total Hamiltonian $H = H_l + H_{\text{int}}$ should be complemented by the interaction of the soft mode coordinate with the elastic strain and the acoustic phonons, which –for transparency – is omitted here. The distortion from the high symmetry phase caused by the phase transition is taken into account by introducing fluctuations about the thermal average value: $Q_j(l) = A_j + r_j(l)$ . From equations 1 and 2 the equations of motion for $Q_j(l)$ and $r_j(l)$ are derived by standard procedures which yield:

$$-\frac{d^2}{dt^2}Q_j(l) = \Omega_0^2 Q_j(l) - \sum_{j'} v_{jj'}(ll')Q_{j'}(l') + \Gamma_1 Q_j^3(l) + \Gamma_2 \sum_{j\neq j'} Q_j(L)Q_{j'}^2(l') \tag{3}$$

and

$$-\frac{d^2}{dt^2}r_j(q) = \Omega_0^2 r_j(q) - \sum_{j'} v_{jj'}(ll')r_{j'}(q) + 3\Gamma_1(A_j^2 + \Delta_{jj'})r_j(q) +$$

$$\Gamma_2 \sum_{j\neq j'} [2(A_j A_{j'} + \Delta_{jj'})r_{j'}(q) + (A_{j'}^2 + \Delta_{jj'})r_j(q)] \tag{4}$$

where $\Delta_{jj'} = \langle r_j(l)r_{j'}(l)\rangle$ is the expectation value of the quadratic term. The equations can be diagonalized through the canonical transformation:

$$r_j(q) = \sum_{j'} b_{jj'}(q)s_{j'}(q), \quad P_j(q) = \sum_{j'} p_{j'}(q)b_{jj'}^{-1}(q) \tag{5}$$

Now the eigenfrequencies $\omega_j$ depend on the correlation function $\Delta_{jj'}$ which is evaluated by using the fluctuation dissipation theorem:

$$\Delta_{jj'} = \frac{1}{N}\sum_{qj} b_{jj}(q)b_{j'j'}(-q)[2\omega_j(q)]^{-1}\coth\tfrac{1}{2}\beta\omega_j(q) \tag{6}$$

and has to be calculated self-consistently. By inspection of equations 2, 3, and 6 it is obvious that the temperature dependence of the renormalized eigenfrequency is solely determined by equation 6, which in the limit $\omega_j(q) < \beta$ leads to a linear T-dependence. A detailed determination of the T-dependence can be obtained by calculating the free energy which has been done by various authors [see, e.g., Ref. 16 and references therein]. The general conclusion is that the soft mode follows a Curie law, namely $\omega_f^2 \approx (T - T_C)$ above $T_C$ and $\omega_f^2 \approx -2(T - T_C)$ below $T_C$. This dependence is indeed seen in many perovskite ferroelectrics, however, only as long as intermediate temperatures are considered. For high temperatures and at low temperatures substantial deviations from this simple law are observed which cannot be accounted for in the rigid ion models, described above. At high temperatures the soft mode starts to saturate with a critical exponent tending to 1/3. At low temperatures and especially in the quantum regime, the soft mode is stabilized by quantum fluctuations which inhibits a real instability and has led to the concept of quantum paraelectricity [13]. Since it has rapidly been realized that the quantum paraelectric regime cannot be adequately described within rigid ion anharmonic models, a phenomenologically derived equation, the Barrett formula [14], is typically applied here. However, a microscopically justified approach was only given later within the polarizability model, first described by Migoni, Bilz and Bäuerle in 1976 [15].

### a) Polarizability effects

The above described model for ferroelectrics relies on the fact that perovskite oxide ferroelectrics are insulators with a large band gap. In this case it should be possible to treat the electrons as being rigidly coupled to their ionic cores and implement the electronic potential in the ionic core potential. However, more than 90% of all ferroelectrics, including hydrogen bonded ones, are oxides [16]. As such, one would expect that the oxygen ion plays a dominant role for the occurrence of ferroelectricity [17]. Indeed, it is known since long that the oxygen ion and also the heavier chalcogenide ions in their doubly negatively charged state are peculiar as compared to most other ions since they are unstable as free ions [18]. This is obvious from the fact that in a solid their polarizabilities are strongly dependent on the disposable volume [19]. In simple oxides like MgO, CaO, SrO, the oxygen ion polarizability $\alpha(O^{2-})$ depends linearly on the volume, in tetrahedrally coordinated oxides this dependence is enhanced to a volume squared dependence. In anisotropic configurations as realized in spinels and ferroelectric oxides $\alpha(O^{2-}) \approx V^{3-4}(O^{2-})$. In addition, it is a strong function of temperature and pressure. These properties are a consequence of the fact that the 2p-electrons tend to delocalize and hybridize with nearest neighbor transition metal d-states. The degree of hybridization can be triggered through the dynamics thereby leading to a *dynamical covalency* [20]. Early on it has been emphasized that the oxygen ion cannot be assigned a Goldstein fixed rigid ion radius but that instead space increments have to be introduced which can be related to p-d hybridization degrees [18].

An understanding of the above described properties can be achieved by calculating quantum mechanically the oxygen ion polarizability [21]. This is possible by stabilizing the free $O^{2-}$ ion by a homogeneously 2+ charged sphere, known as the Watson model [22]. Upon varying the radius $R_W$ of this sphere, the oxygen ion polarizability can be obtained as a function of its volume. The Watson sphere describes effectively the crystalline surrounding of the oxygen ion with the limitation that crystalline Coulomb potential is homogeneous. In addition, certain conditions for the Watson potential $V_W$ (figure 1) have to be fulfilled, in order to be compatible with a homogeneous crystalline surrounding:

a)  the depths of the potential well equals the Madelung potential at the oxygen ion lattice site and thus defines the Watson radius which implies that the Watson radius $R_W$ mimics the lattice constant;

b)  the treatment of the unstable ion problem by an atomic model requires an asymptotic compensation of the Coulomb potential $V_C$ by the Watson potential $V_W$ for large radii.

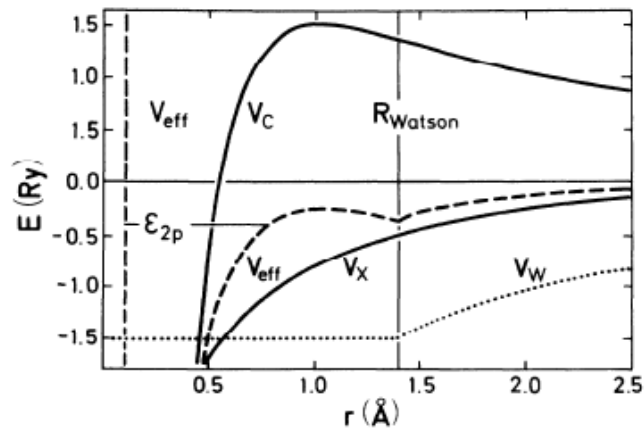

**Figure 1** The Watson sphere model of the $O^{2-}$ ion: Coulomb potential $V_C$, exchange potential $V_x$, Watson potential $V_W$, resulting potential $V_{eff}$, and the energy level of the 2p electrons $\varepsilon_{2p}$ as a function of the oxygen ion radius $r$ [after Ref. 17].

By varying $R_W$, the Watson sphere radius, the wave functions and charge densities of the oxygen ion can be calculated for different crystalline surroundings. These wave functions are used to compute the oxygen ion polarizability within the formalism developed by Thorhallsson et al. [23] who approximated the weak electric field disturbed wave functions by a variational procedure including perturbation theory. In a weak field the perturbed wave function $\Phi$ is represented by: $\Phi = \Phi_0(1 + \omega)$, where $\Phi_0$ represents the unperturbed wave function with energy $E_0$. For $\omega$ various approaches have been made, however, satisfactory agreement with experiment was achieved by the form $\omega = F(\mu r + \nu r^2)\cos\theta$ [23], with $r$ denoting the position of the electron under consideration, $F$ is a uniform electric field, and $\nu, u$ are the potential due to the field at the position $r$ and the solution of a second order differential equation. The energy in the presence of a weak field is given by:

$$E = E_0 + \tfrac{1}{2}[A_0\mu^2 + 2A_2\nu^2 + \tfrac{4}{3}(B_0\mu + 2A_1\mu\nu + B_1\nu)]F^2 \qquad (7)$$

and

$$A_k = \left\langle \Phi_0 \mid \sum_i r_i^k \mid \Phi_0 \right\rangle \qquad (8)$$

$$B_k = \left\langle \Phi_0 \mid \sum_{i,j} r_i^k (\vec{r}_i \vec{r}_j) \mid \Phi_0 \right\rangle \qquad (9)$$

Minimization of $E$ with respect to $\mu, \nu$ yields:

$$\mu = (4A_1B_1 - 6A_2B_0)/(9A_0A_2 - 8A_1^2) \qquad (10)$$

$$\nu = (4A_1B_0 - 3A_0B_1)/(9A_0A_2 - 8A_1^2) \qquad (11)$$

Upon assuming that $E = E_0 - 1/2\alpha F^2$, the dipole polarizability becomes:

$$\alpha = -[A_0\mu^2 + 2A_2\nu^2 + \tfrac{4}{3}(B_0\mu + 2A_1\mu\nu + B_1\nu)] \qquad (12)$$

The calculated wave functions of the $O^{2-}$ ion are shown in figure 2 (a) as a function of $R_W$.

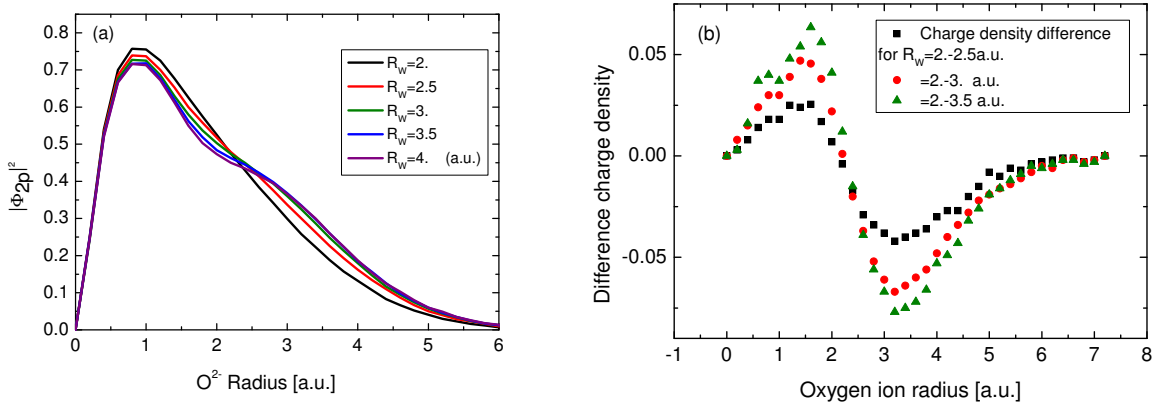

**Figure 2** (a) The oxygen ion 2p wave function as a function of the $O^{2-}$ radius for different Watson sphere radii as indicated in the figure.
(b) Difference charge density as a function of the $O^{2-}$ radius for different Watson sphere radii as indicated in the figure.

Obviously, the $O^{2-}$ wave function does not undergo a rigid shift with increasing $R_W$, but develops a shoulder to interstitial regions as is expected for charge transfer. This can be seen more clearly by inspection of the difference charge density (figure 2 (b)) which reveals the clear tendency of forming space increments as originally proposed by Biltz and Klemm [18]. The polarizability $\alpha(O^{2-})$ as a function of $R_W$ is shown in figure 3 in comparison to the one of the isoelectronic $F^-$ ion.

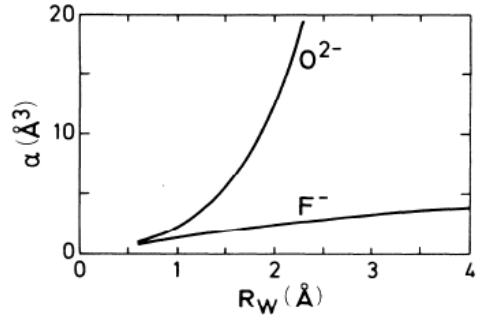

**Figure 3** The polarizabilities α of $O^{2-}$ and $F^-$ as functions of $R_W$ [after Ref. 17].

While $\alpha(F^-)$ rapidly converges to its rigid ion value, $\alpha(O^{2-})$ diverges with increasing $R_W$ and is proportional to $R_W{}^3$ in the physically relevant regime.
In order to evaluate wave functions and polarizabilities of anisotropic and covalent oxides, the Watson sphere model has to be extended [21]. Anisotropy can be incorporated by considering an ellipsoidal surrounding of the oxygen ion which can be simulated by taking a weighted average over Watson spheres with different radii $R_{W1}$, $R_{W2}$. Using the closure approximation and a rotational ellipsoidal polarizability, $\alpha(O^{2-})$ experiences an additional volume dependence as compared to the isotropic case. In covalent oxides the oxygen 2p orbital can be described by:

$$\Phi_{2p} = \lambda(r)e_{3d} + [1 - \lambda(r)]e_{2p} \qquad (13)$$

where $\lambda(r)$ corresponds to the distance $r$ dependent admixture of the transition metal d-state. Not only change cell volume variations this admixture but also can lattice displacements from phonons contribute to it.
Interesting results for the $O^{2-}$ ion have been obtained by Prat [24] by calculating the energy levels and total energies for different configurations and symmetries of the oxygen ion within the unrestricted Hartree Fock theory. He found that the spherically symmetrical solution of the wave functions produces an unstable state, whereas an ellipsoidal charge distribution leads to a metastable state. This fact is in agreement with chemical structure data where mainly the two and fourfold coordination of oxygen in solids are known emphasizing the in-out tendency of p-electrons pairs with opposite spins.

### b) The lattice dynamical model

The failure of rigid ion models in accounting for the dynamics of perovskite oxides and their temperature dependent characteristics has been addressed early by Cochran and coworkers [25] and later by Cowley [26] and Stirling [27]. In order to solve these problems they introduced a shell model description of $ABO_3$ systems, where each ion is surrounded by its own electronic shell which is coupled to the shells of the nearest neighbors. In this approach all shells are isotropically polarizable with shell charges $Y_i$ (i=A, B, O) and core shell coupling constants $K_i$. In order to realistically model ferroelectric perovskites with the aim to reproduce experimental data and be predictive with respect to novel systems and new emergent physics, the model has to be extended to account for nonlinear oxygen ion polarizability effects as emphasized in the preceding paragraph.

Here it is shown, that an analytically tractable form of the 3D model [15] can be obtained by reducing the systems dimensionality and thereby also reducing substantially the involved model parameters [17, 28]. As outlined above, essential to the physics of perovskites is the nonlinear onsite and directional polarizability of the oxygen ion which acts essentially in the hybridization with the central transition metal ion through dynamical covalency. The importance of the p-d hybridization has been demonstrated for mixed crystals of $KTa_{1-x}Nb_xO_3$ [29], where hyper Raman data could successfully be modeled within the polarizability model. The importance of that work for the following lies in the fact that for the understanding of the data a single quantity turned out to be the decisive one, namely the effective core-shell coupling $K_{OB}(T)$ [30]. This finding implies that only a limited number of parameters is relevant for the description of the dynamical and temperature dependent properties of ferroelectric perovskites. On this basis the above described modeling can be reduced to a simpler and more transparent approach by considering the dynamics only along one specific direction which corresponds to the projection onto the axis where the ferroelectric polarization develops. Simultaneously, the $BO_3$ cluster is replaced by one polarizable mass $M_1$ with core shell couplings constants $g_2$ and $g_4$. While the former coupling includes the attractive Coulomb interactions and causes the lattice instability, the latter is a stabilizing fourth order attractive term which is directly related to $K_{OB,B}$. The second sublattice mass is a rigid ion mass $M_2$ referring to the A sublattice. The three-dimensional character, important for critical phenomena is, however, preserved in the phase-space integrations. The Hamiltonian of this model is given by: $H = T + V$ with

$$T = \tfrac{1}{2}\sum_n (M_1\dot{u}_{1n}^2 + M_2\dot{u}_{2n}^2 + m_{e1}\dot{v}_{1n}^2) \tag{14a}$$

$$V = \frac{1}{2}\sum_n \left[ f'(u_{1n} - u_{1n-1})^2 + f(u_{2n} - v_{1n})^2 + f(u_{2n+1} - v_{1n})^2 + g_2(v_{1n} - u_{1n})^2 + \tfrac{1}{2}g_4(v_{1n} - u_{1n})^4 \right] \tag{14b}$$

where $u_{in}, v_{1n}$ are the core and shell displacements of ion i (i=1, 2) and shell in cell $n$

$m_{e1}$ being the shell mass of ion $M_1$. $f', f$ are second and nearest neighbor harmonic core-

core and core-shell coupling constants. The onsite potential in the core shell interaction is of double-well character since $g_2$ is attractive, and can be written as: $V(w_{1n}^2) = \frac{1}{2} g_2 w_{1n}^2 + \frac{1}{4} g_4 w_{1n}^4$ using the definition $w_{1n} = u_{1n} - v_{1n}$. The new relative displacement coordinate $w_{1n}$ characterizes the polarization and has the limits of a fully delocalized shell for $w_{1n} \to \infty$, i.e., ionization, and a rigidly bound shell for $w_{1n} \to 0$. By applying the adiabatic condition: $\delta V / \delta v_{1n} = 0$ the equations of motion become:

$$M_1 \ddot{u}_{1n} = f'(u_{1n+1} + u_{1n-1} - 2u_{1n}) + g_2 w_{1n} + g_4 w_{1n}^3 \tag{15a}$$

$$M_2 \ddot{u}_{2n} = f(v_{1n+1} + v_{1n} - 2u_{2n}) \tag{15b}$$

$$0 = -g_2 w_{1n} - g_4 w_{1n}^3 + f(u_{2n} + u_{2n-1} - 2v_{1n}) \tag{15c}$$

Obviously, the nonlinear term proportional to $w_{1n}^3$ does not admit for a straightforward solution of the coupled equations. However, interesting novel solutions are obtained from it which will be discussed in detail below. In order to find plane-wave solutions of the above equations, the self-consistent phonon approximation (SPA) is used which corresponds to a cumulant expansion of the nonlinear terms, namely:

$g_4 w_{1n}^3 \cong 3 g_4 w_{1n} \left\langle w_{1n}^2 \right\rangle_T$, where

$$\left\langle w_{1n}^2 \right\rangle_T = \sum_{qj} \frac{\hbar}{N \omega_{qj}} w_1^2(qj) \coth \frac{\hbar \omega_{qj}}{2 k_B T} \tag{16}$$

contains all the dynamical information through the SPA eigenvalues $\omega_{qj}$ and related eigenvector $w_1(qj)$ for all phonon branches $j$ and all wave vectors $q$ in the first Brillouin zone. With the definition $g_T = g_2 + 3 g_4 \left\langle w_{1n}^2 \right\rangle_T$ the equations of motion yield stable and pseudo-harmonic solutions which admit to solve the dynamical matrix in $q$-space:

$$\omega^2 \begin{vmatrix} M_1 & 0 \\ 0 & M_2 \end{vmatrix} = D(q) \begin{vmatrix} U_1 \\ U_2 \end{vmatrix} \tag{17}$$

with the force constant matrix given by:

$$D(q) = 2\tilde{f} \begin{vmatrix} 1 + A_1 \sin^2 qa & -\cos qa \\ -\cos qa & 1 + A_2 \sin^2 qa \end{vmatrix} \tag{18}$$

and $2\tilde{f} = 2fg_T / (2f + g_T)$, $A_1 = 2f' / \tilde{f}$, $A_2 = 2f / g_T$. The core-shell eigenvector $W_1$ is related to the eigenvectors $U_1, U_2$ though the equation:

$$W_1 = -2\tilde{f} / g_T U_1 + 2\tilde{f} / g_T \cos(qa) U_2 . \tag{19}$$

The dispersion relations are now given by:

$$\omega_\pm^2(q) = \tilde{f} \left[ \frac{1}{M_1} (1 + A_a \sin^2 qa) + \frac{1}{M_2}(1 + A_2 \sin^2 qa) \right] \pm$$

$$\tilde{f} \left\{ \left[ \frac{1}{M_1}(1 + A_a \sin^2 qa) - \frac{1}{M_2}(1 + A_2 \sin^2 qa) \right]^2 + \frac{4 \cos^2 qa}{M_1 M_2} \right\}^{1/2} \tag{20}$$

In the long wave length limit the two modes approach:

$$\omega_+^2 = \frac{2\tilde{f}}{\mu} \ (\mu \text{ being the reduced cell mass}) \tag{21a}$$

$$\omega_-^2 \cong \frac{2}{M_1 M_2}(f + 2f' \sin^2 qa) \tag{21b}$$

At the zone boundaries the corresponding values are:

$$\omega_+^2 = \frac{2f}{M_2} \tag{22a}$$

$$\omega_-^2 = \frac{1}{M_1}(2\tilde{f} + 4f') \tag{22b}$$

From these results it is obvious that the zone center optic mode, i.e., the soft mode, is strongly temperature dependent through its dependence on $g_T$, whereas the acoustic mode is T-independent in the limit $q \approx 0$. The opposite happens at the zone boundary where the optic mode adopts the rigid ion value while the acoustic mode depends on T through $g_T$. The picture is mixed for small momentum where strong mode-mode coupling sets in, which causes pronounced anomalies in the acoustic mode. These have been shown to appear as precursor effects before the actual structural instability and signal order / disorder dynamics in addition to the long wave length displacive dynamics [31, 32]. Note, however, that both dynamics are characterized by very different length scales.

The temperature dependence of $\omega_+^2(q=0) = \omega_F^2$ is given by a single self-consistent equation:

$$\mu\omega_F^2 = 2f(g_2 + 3g_4 I_F(T)/[2f + g_2 + 3g_4 I_F(T)] \tag{23}$$

where $I_F(T)$ is an integral over the 3D Brillouin zone which contains all dynamical information. In order to extract the analytical behavior of the soft mode over a wide temperature regime, an interpolation scheme between acoustic and optic mode can be used where an isotropic dispersion is assumed:

$$\omega_f^2(q) = \frac{2\tilde{f}}{\mu} + \frac{4f'}{M_1}a^2 q^2 \tag{24}$$

from which the relation of the core eigenvectors is obtained:

$$M_2 U_2 = -M_1 U_1 \cos qa \tag{25}$$

Together with the normalization condition $M_1 U_1^2 + M_2 U_2^2 = 1$ the q-dependent relative core shell displacement eigenvector which enters equation 23 is obtained:

$$W_1^2(q) = \frac{1}{M_1}\left[\frac{2\tilde{f}}{g}\right]^2\left[1 + \frac{M_1}{M_2}\cos^2 qa\right] \tag{26}$$

While this approximation is quite useful in evaluating analytically the soft mode temperature dependence, it is insufficient with respect to mass dependencies of the soft mode, the phase transition temperature and certain dynamical regimes. In order to work these effects out in detail, the above equation has to be replaced by the exact one:

$$W_1^2(q) = \sum_{q,\omega} \frac{4\tilde{f}^2 \cos^2 qa (M_1\omega^2 - 4f'\sin^2 qa)^2}{M_2 g_T^2 (M_1\omega^2 - 4f'\sin^2 qa - 2\tilde{f})^2 + M_1 4\tilde{f}^2 g_T^2 \cos^2 qa} \tag{27}$$

which explicitly includes the sums over momenta and frequencies. Note, that in the self-consistent calculations of temperature dependent quantities the latter equation 27 is always used instead of the approximate equation 26.

However, based on the approximate equation 26, the integral $I_F(T)$ is easier to evaluate and given by:

$$I_F(\omega_F, T) \approx \frac{3\hbar}{\mu\omega_D^3} \int_0^{\omega_D} \frac{\omega^3}{(\omega_F^2 + \omega^2)^{1/2}} \coth\left[\frac{\hbar(\omega_F^2 + \omega^2)^{1/2}}{2k_B T}\right] d\omega \tag{28}$$

where $\omega_D = \sqrt{f_D/\mu}$ is the Debye frequency and $f_D$ the related force constant. The analytical behavior of the integral controls the temperature dependence of the soft mode in different regions.

In the high temperature $2k_B T > \hbar\omega_D$ the hyperbolic cotangent can be replaced by its reciprocal argument:

$$I_F(\omega_F, T) \approx \frac{3k_B T}{\mu\omega_D^2}\left[1 - \frac{\omega_F}{\omega_D}\arctan\left(\frac{\omega_D}{\omega_F}\right)\right] \tag{29}$$

from which the mean-field relation is derived for $\omega_F << \omega_D$:

$$\omega_F^2 = \frac{|g_2|}{\mu T_C}(T - T_C), \quad (T \geq T_C) \tag{30}$$

and $T_C = f_D |g_2|/(9g_4)$ which is mass independent. Indeed, experimentally $T_C$ is independent of the mass as long as $T_C$ is finite and rather large. With decreasing $T_C$ and approaching the quantum limit this is no longer correct and a pronounced mass dependence of the soft mode appears [33] which will be discussed in the subsequent chapter 5.

Upon including in the integral, both, the acoustic and the optic mode instead of the interpolation scheme, the high temperature limit of the soft mode differs from the above mean-field result, since a saturation of the soft mode sets in, approximately given by:

$$\omega_F^2 \approx \frac{2f}{M_2}\left[1 - \left(\frac{f_D^2}{3g_4 k_B T}\right)^{1/3}\right] \tag{31}$$

At very high temperature the saturation of the soft mode is complete and it approaches the value $\omega_F^2 \to 2f/\mu$. Note, that the saturation exponent differs from 1/3 when non-adiabatic effects are included where it becomes 1/2 instead [30, 34, 35].

In the classical and quantum regime and for $\omega_F = 0$, T=$T_C$, analytical expressions for the integral can be derived where three cases need to be distinguished:

1. $2k_B T/\hbar \geq \omega_D >> \omega_F$                                      (32a)

2. $\omega_D >> 2k_B T/\hbar > \omega_F$.                                      (32b)

3. $\omega_D >> \omega_F > 2k_B T/\hbar$                                        (32c)

While case 1. applies to classical ferroelectrics, case 2. corresponds to the quantum regime, and case 3. to incipient ferroelectrics. For these three cases the integral adopts the following forms:

$$1. \quad I_F \approx \frac{3k_B T}{f_D}\left[1 - \frac{\pi\omega_F}{2\omega_D} + \left(\frac{\omega_F}{\omega_D}\right)^2\right] \tag{33a}$$

$$2. \quad I_F \approx \frac{3\hbar\omega_D}{4f_D}\left[1 + \left(\frac{2k_B T}{\hbar\omega_F}\right)^2 - \frac{\pi\omega_F}{\omega_D}\frac{2k_B T}{\hbar\omega D} + \left(\frac{\omega_F}{\omega_D}\right)^2\left(1 + \ln\left(\frac{2k_B T}{\hbar\omega D}\right)\right)\right] \tag{33b}$$

$$3. \quad I_F \approx \frac{3\hbar\omega_D}{4f_D}\left[1 + \left(\frac{\omega_F}{\omega_D}\right)^2\left(\frac{1}{2} + \ln\left(\frac{\omega_F}{2\omega_D}\right)\right)\right] \tag{33c}$$

While for the incipient ferroelectrics the T-dependence has vanished, the two other cases still retain it, as expected. The transition temperature for the first case is the one given by equation 30, whereas in the quantum region the approximate solution for $T_C$ is:

$$k_B T_C \approx -\frac{1}{2}\hbar\left(\frac{f_D}{\mu}\right)^{1/2}/\ln\left(\frac{\mu}{\mu_c}-1\right) \tag{34}$$

with $\mu_c = (9g_4\hbar/4g_2)^2/f_D$.

From the above it can be concluded that in most cases the mean-field behavior is observed. For $\omega_F \to 0$ the critical exponent is larger than 1 and approaches 2, which mostly occurs in the quantum limit. There is – of course – a crossover regime between both exponents which has been observed in KTaO$_3$ [30], where the dimensionality of the system changes according to renormalization group theory [36]. This theory also predicts that in the quantum regime logarithmic corrections in the soft mode temperature dependence appear.

While until now only the paraelectric case has been discussed, in the following the behavior in the ferroelectric region will be considered. This can be carried through by considering the free energy of the system as being composed of a sum of contributions which stem from the rigid ion potential $V_0$, the polarizability potential $V_w$, and the temperature dependent dynamical part due to SPA phonons:

$$F = V_0 + V_w + k_B T\sum_q\ln\left[2\sinh\left(\frac{\hbar\omega_F(q)}{2k_B T}\right)\right] \tag{35}$$

with $V_w = \frac{1}{2}g_2\langle w_1\rangle_T^2 + \frac{1}{4}g_4\langle w_1\rangle_T^4$. The equilibrium configuration is given by:

$$\frac{\partial F}{\partial\langle w_1\rangle_T} = 0 = \langle w_1\rangle_T\left(g_2 + g_4\langle w_1\rangle_T^2\right) + \frac{1}{2N}\sum_q\frac{\hbar}{2\omega_F(q)}\frac{\partial\omega_F^2(q)}{\partial\langle w_1\rangle}\coth\left(\frac{\hbar\omega_F(q)}{2k_B T}\right) \tag{36}$$

Here $\omega_F$ is given by the interpolating form equation 24. Upon replacing the differentiation of $\omega_F$ with respect to the polarizability coordinate by a differential with respect to $g_T$, the sum in equation 36 becomes

$$\frac{\partial \omega_F^2(q)}{\partial g_T} = \frac{\partial \tilde{f}}{\partial g_T}\left[\frac{2}{\mu} - \frac{2}{M_2}\sin^2 qa\right] = \frac{1}{M_1}\left(\frac{2\tilde{f}}{g_T}\right)\left(1 + \frac{M_1}{M_2}\cos^2 qa\right) = w_1^2(q) \tag{37}$$

From this derivation the free energy adopts the following form:

$$\frac{\partial F}{\partial \langle w_1 \rangle_T} = \langle w_1 \rangle_T [g_2 + g_4 \langle w_1 \rangle_T^2 + 3g_4 \langle w_1^2 \rangle_T] \tag{38}$$

which is zero for $\langle w_1 \rangle_T = 0$ (paraelectric phase) or:

$$\langle w_1 \rangle_T = \pm\left[\frac{g_2}{g_4} - 3\langle w_1^2 \rangle_T\right]^{1/2} = \pm[2(\langle w_1^2 \rangle_{T_c} - \langle w_1^2 \rangle_T)]^{1/2} \tag{39}$$

in the ferroelectric phase. This implies that in the ferroelectric phase the core shell force constant is given by:

$$g_{T,FE} = -2g_2 - 6g_4 \langle w_1 \rangle_T^2 = -2g_{T,PE} \tag{40}$$

which is analogous to the one obtained from a Landau free energy functional [16] and consequently the slope of $\omega_F^2$ is steeper in the ferroelectric region as compared to the paraelectric phase.

In this chapter the polarizability model has been reviewed. It has been shown that the important ingredients of perovskite ferroelectrics are incorporated in this approach where especially various temperature regimes for the soft mode are predicted including saturation at high temperatures, a mean-field region at intermediate temperatures, the suppression of ferroelectricity due to quantum fluctuations at low temperatures, and the temperature independency of the soft mode in incipient ferroelectrics. The model parameters are considerably reduced as compared to the 3D version [15] which contributes significantly to an increased transparency. All parameters are directly related to experimental data, with the exception that the double-well defining quantities have to be derived self-consistently. As will be shown in subsequent chapters, various perovskite ferroelectrics are well modeled within this approach and excellent agreement with experimental data is obtained.

Besides of reproducing the overall temperature dependent characteristic properties of these compounds in excellent agreement with experimental data, the model exhibits various new nonlinear and local solutions which are of importance for understanding order / disorder phenomena [31, 32] together with relaxor properties [37 – 39], to be discussed later on where it will be shown that crossover between different dynamics and displacive – relaxor behavior exist.

## 3. Applications

As mentioned in the introduction the ferroelectric properties of perovskite oxides have first been discovered in 1945 and the search for novel materials is still going on. Since these systems are very stable, have good growing conditions and high dielectric constants, a large field of applications is based on them. Simultaneously the rather simple structure of $ABO_3$ compounds is a challenging tool for theoretical investigations. As has first been pointed out by Cochran this structure has some features in common with the cubic CsCl structure [25]. Therefore, the reduction from 3D to pseudo 1D can rather straightforwardly been carried through. Within the pseudo 1D model three perovskite crystals have been analyzed in deeper detail, two of which are incipient ferroelectrics, $SrTiO_3$ and $KTaO_3$, and one with classical soft mode behavior and high $T_C$, $PbTiO_3$. For all three compounds the momentum dependent lowest transverse acoustic and optic modes have been calculated, and the temperature dependence of the soft mode has been derived self-consistently within the above outlined scheme. While for $SrTiO_3$ and $KTaO_3$ the A sublattice in $ABO_3$ is a rigid ion, this is no longer true for Pb in $PbTiO_3$ which carries its own polarizability. For this compound an additional core-shell coupling $g_2^{(2)}$ has to be included together with a core-core coupling $f_2'$ between the Pb sublattice. The coupling constants entering the theoretical analysis are fixed through the experimental data, and the self-consistently derived double-well potential defining parameters are optimized to best describe the soft mode temperature dependence.

In figures 4(a – c) the theoretical [40] and experimental data are compared to each other. Overall good agreement with the experiment is achieved for $KTaO_3$ [41 – 43] and $PbTiO_3$ [44]. For $SrTiO_3$ [45 – 47] the in between lying optic modes have been ignored for symmetry reason. Instead, the long wave length soft mode and the high energy zone boundary optic mode have been interpolated thereby ignoring the crossing in between lying modes. This approximation is justified by experiment since the in between lying branches are of different symmetry. An additional support for this choice is obtained from the temperature dependence of the squared soft mode frequency where the calculations are in excellent agreement with the experimental data.

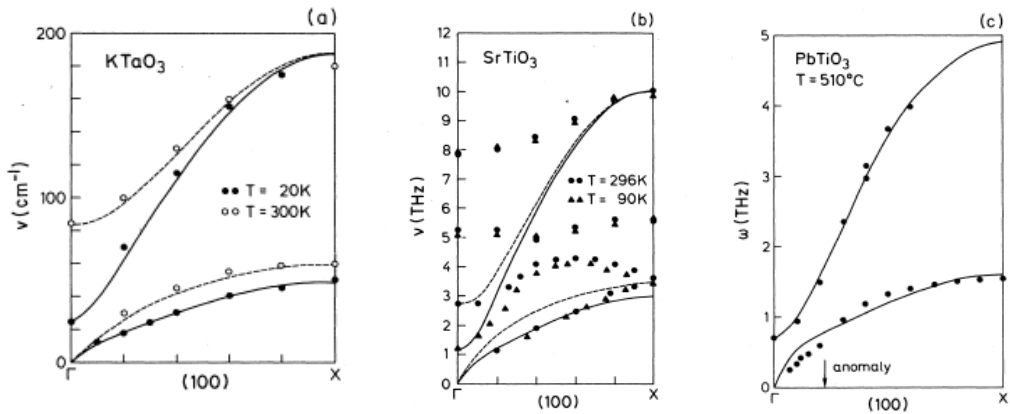

**Figure 4** Comparison of experimental and theoretical dispersion curves for (a) $KTaO_3$ at 20K (solid circles and full lines) and at 300K (open circles and dashed lines) [41 – 43]. (b) $SrTiO_3$ at 90K (triangles and full lines) and at 296K (full circles dashed lines) [45 – 47]; (c) $PbTiO_3$ at 510C (circles experimental data from Ref. 44, full lines theory) [after Ref. 40].

The dispersion curves of all three compounds exhibit an anomaly in terms of a dip at small wave vector, which will be discussed in more detail below. This anomaly gets more and more pronounced the closer the temperature approaches $T_C$ in $PbTiO_3$, respectively $T \rightarrow 0$ in $SrTiO_3$ and $KTaO_3$.

The temperature dependencies of the squared soft mode frequencies are displayed in figures 5 (a – c) again in comparison to experimental data.

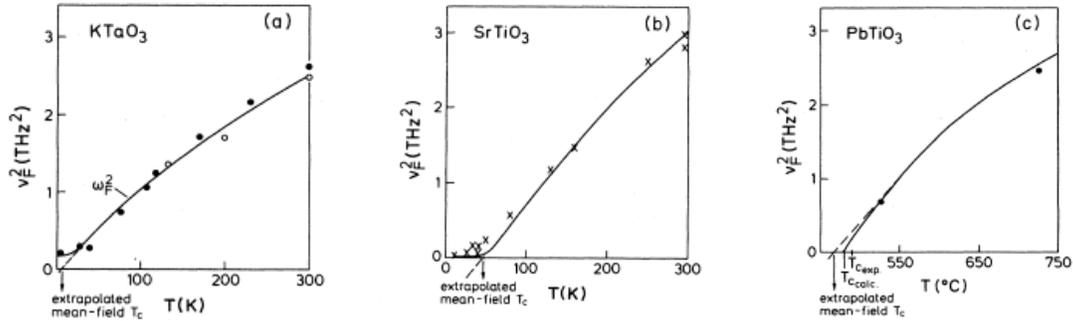

**Figure 5** Comparison of experimental (circles [41 – 43], crosses [45 – 47] and circles [44] and theoretical temperature dependencies (full lines) of $\omega_F^2$ for (a) KTaO$_3$, (b) SrTiO$_3$, (c) PbTiO$_3$ [after Ref. 40].

While over a large temperature regime the soft modes follow mean-field behavior, namely $\omega_F^2 \approx (T - T_c)^\gamma$, $\gamma = 1$, deviations from this are apparent at low temperatures for SrTiO$_3$ and KTaO$_3$, and close to T$_c$ for PbTiO$_3$. While in the former systems quantum fluctuations cause an enhancement of the dimensionality [36] and induce these deviations, in PbTiO$_3$ the first order nature of the phase transition is responsible for the observed deviation from mean-field behavior. A more detailed discussion of the quantum paraelectric compounds follows in a subsequent chapter on isotope induced ferroelectricity. The model parameters for all three systems are given in table 1 [40].

The potentials of the three systems are compared to each other in figure 6. While the ones of SrTiO$_3$ and KTaO$_3$ are very shallow, the one of PbTiO$_3$ is deep and narrow consistent with the vanishing T$_C$ in the former compounds and the high T$_C$ of the latter. However, opposite to e.g. EuTiO$_3$ [48, 49] the potential of PbTiO$_3$ has almost the same width as the one of the two quantum paraelectrics, supporting displacive dynamics.

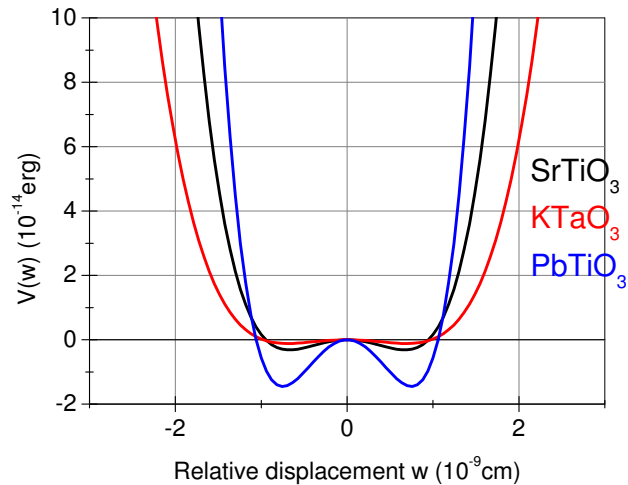

**Figure 6** The double-well potentials as a function of the polarizability coordinate $w$ for SrTiO$_3$ (black), KTaO$_3$ (red), and PbTiO$_3$ (blue).

**Table 1** Parameters used to calculate the dispersions and temperature dependencies of the soft modes of SrTiO$_3$, KTaO$_3$, PbTiO$_3$.

| | SrTiO$_3$ | KTaO$_3$ | PbTiO$_3$ |
|---|---|---|---|
| $M_1$ $(10^{-22}$g$)$ | 1.549 | 2.88 | 1.65 |
| $M_2$ $(10^{-22}$g$)$ | 1.461 | 0.649 | 3.44 |
| $f$ $(10^4$g s$^{-2})$ | 14.405 | 4.067 | 13.00 |
| $f'$ $(10^4$g s$^{-2})$ | 1.268 | 0.581 | 0.912 |
| $f'_2$ $(10^4$g s$^{-2})$ | | | 17.27 |
| $g_2$ $(10^4$g s$^{-2})$ | | | 28.00 |
| $g_2^{(2)}(10^4$g s$^{-2})$ | -1.41 | -0.49 | -5.171 |
| $g_4$ $(10^{22}$gs$^{-2}$cm$^2)$ | 1.57 | 0.51 | 4.606 |

Applications of the model to other ferroelectrics has been done as well [40], however, since these are not perovskites further details are omitted here.

BaTiO$_3$ is the classical ferroelectric perovskite. For this compound seemingly controversial results about its transition dynamics have been reported [59 – 61]. From light and inelastic neutron scattering techniques it has been concluded that the dynamics are in the displacive limit [53]. On the contrary, local probe testing experiments like NMR and EPR have been interpreted in terms of an order/disorder phase transition [54 – 57]. In addition, it has been shown by EXAFS techniques [58] that the local structure already far above the real lattice instability deviates from cubic symmetry suggesting precursor dynamics [59 – 61]. The interpretation of these results [31, 38] will be given in chapter 7 where it will be shown that crossovers between the various transitional aspects coexist and naturally lead to precursor effects.

## 4. Extensions of the model

### a) Travelling wave solutions

The pseudo 1D model can be investigated for exact solutions by studying it beyond the SPA [62]. The solvability of the equations of motion is essentially more complicated as compared to $\Phi_4$-models since general transformation methods are not applicable [63 – 65]. However, by using the continuum limit, exact travelling wave solutions exist. Besides of kinks, which are known from $\Phi_4$-models, also periodons, extended breather solutions and travelling pulses are found which are absent in other nonlinear Hamiltonians. The classification and analysis of travelling wave solutions in the continuum limit can be used to study the corresponding solutions in the lattice case, where they are likely to exist as well. Whether they remain stable in that case is an open issue.

For convenience the equations of motion in the adiabatic approximation are repeated below:

$$M_1 \ddot{u}_{1n} = f'(u_{1n+1} + u_{1n-1} - 2u_{1n}) + g_2 w_{1n} + g_4 w_{1n}^3 \tag{41a}$$

$$M_2 \ddot{u}_{2n} = f(v_{1n+1} + v_{1n} - 2u_{2n}) \tag{41b}$$

$$0 = -g_2 w_{1n} - g_4 w_{1n}^3 + f(u_{2n} + u_{2n-1} - 2v_{1n}) \tag{41c}$$

Solutions to equations 41 in the travelling wave form are obtained with:

$$u_{1n}(t + 2a/v) = u_{1n}(t) \tag{42a}$$

$$u_{2n}(t + 2a/v) = u_{2n}(t) \tag{42b}$$

$$w_{1n}(t + 2a/v) = w_{1n}(t) \tag{42c}$$

where $a$ is the interatomic distance and $v$ the phase velocity. In the continuum limit an expansion with respect to $\tau = 2a/v$ is used:

$$u_{1n\pm1}(t) = u_{1n}(t) \pm \tau \dot{u}_{1n}(t) + \frac{1}{2}\tau^2 \ddot{u}_{1n}(t) \tag{43}$$

with the corresponding replacements for $u_{2n\pm1}(t), w_{1n\pm1}(t)$. This admits to eliminate the core displacement coordinates and results in a single equation for the polarizability displacement at arbitrary lattice site $n$ such that $w_{1n} \equiv w$:

$$\ddot{w}(1 + \beta - 3\beta w^2/w_F^2) - 6\beta w \dot{w}^2/w_F^2 + w(1 - w^2/w_F^2)g_2/M \tag{44}$$

with the following definitions:

$$w_F^2 = -g_2/g_4, \beta = g_2[1/(1 - v_2^2/v^2)]/2f, \tag{45}$$

$$1/M = [1/\{M_1\{(1 - v_1^2/v^2)\}\}] + [1/\{M_2\{(1 - v_2^2/v^2)\}\}] \tag{46}$$

$$v_1 = a(4f'/M_1)^{1/2}, v_2 = a(2f/M_2)^{1/2} \tag{47}$$

Equation 44 is of Bernoulli type and can be integrated by setting $q(w) = \dot{w}^2$. The first integral of motion defines the effective potential:

$$U(w) = \frac{1}{2}\frac{\mu}{M}\frac{V(w^2) - V(w_0^2)}{(1 + \beta - 3\beta w^2/w_F^2)^2} \tag{48}$$

where $\pm w_0$ are the turning points and

$$V(w^2) = g_2 w^2[1 + \beta - \frac{1}{2}(1 + 4\beta)w^2/w_F^2 + \beta w^4/w_F^4] \tag{49}$$

The effective potential can be solved with respect to time to yield:

$$t = \frac{1}{2}(-M)^{1/2}\int^{w^2}dx\frac{1 + \beta - 3\beta x/w_F^2}{[x(x - w_0^2)R(x)]} \tag{50}$$

$$R(x) = \beta g_4^2 x^2/g_2 + \frac{1}{2}g_4(1 + 4\beta)(x + w_0^2) + g_2(1 + \beta)$$

The solutions to equation 50 have periodic as well as non-periodic character. This differentiation depends on whether $(x - w_0^2)R(x)$ and $-M$ have the same sign for the former type or a non-integrable pole occurs for the latter type.

Periodic type solutions with oscillations across the origin form a continuous set with respect to $v$ and are travelling periodic waves of wave vector $k$:

$$\Omega\tau = 2ak = \left[\frac{-2g_2 2ff'}{f+2f'}\right]^{1/2}\left[1-\frac{w_0^2}{w_F^2}\right]^{1/2} \tag{51}$$

and $k$-dependent amplitude:

$$w_0^2 = w_F^2\left[1+\frac{2ff'}{g_2(f+2f')}a^2k^2\right]. \tag{52}$$

From these solutions the static periodic waves can be recovered where periods 1 to 6 have been obtained in Ref. 66. The period 1 solution corresponds to the ferroelectric case whereas period 2 is the antiferroelectric ground state. For periods N≥3 commensurate periodic waves are obtained where the case N=3 will be discussed below.

For non-periodic solutions two distinct classes have to be differentiated: ferroelectric solutions with asymptotic extrema at $w=\pm w_F$, and paraelectric solutions with $w=0$.

In the ferroelectric case static and slowly propagating solutions of kink type character occur which in the static limit correspond to domain walls. The condition for their existence is given by the velocity limit that $v^2 < v_2^2/(1+g_2/f)$. In the case that real turning points exist, pulse solutions with amplitudes starting from and arriving at the ground state polarization are possible. Upon increasing $w^2$ largely, i.e., the electronic cloud is far away from its associated core, an electronic excitation takes place which can be identified with travelling excitons. Their velocity is, however, limited since the larger the excitation is the slower is their motion. Local ionization processes can no more move through the lattice. The effective potentials are displayed in figures 7 where also paraelectric solutions are included.

Large velocity solutions have either kink or pulse exciton character. They correspond formally to the same solutions as discussed above with requirement that their velocity is increased and the mass ratio $M_2/M_1$ exceeds a critical value.

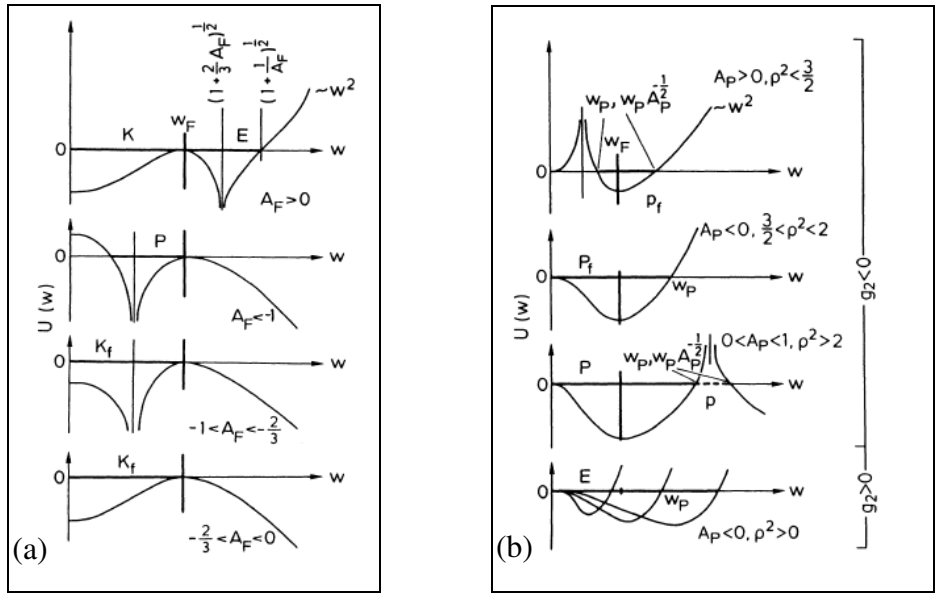

**Figure 7** Effective potentials for ferroelectric (a) and paraelectric (b) waves. K: slow kinks, E: pulse excitons, P: pulses, $P_f$: fast pulses, $p_f$: fast periodic waves; p: slow periodic waves, $A_F$ and $A_P$ are given in Ref. 62.

In the paraelectric phase $R(x)$ has a vanishing root for $R(0) = 0$. The associated effective potential is included in figure 7b. As can be seen there the potential admits for oscillating solutions in addition to paraelectric pulses. Again small and large velocity solutions can be differentiated where the slow ones are large amplitude waves whereas the fast ones are small amplitude waves. For pulse solutions with $w_p^2 / w_F^2 = -2(1 + \beta) / 3\beta$ the pulse is fast with well defined shape (figures 8). For positive values the pulse width increases with increasing $w_p^2 / w_F^2$ to become infinitely wide and finally decay into a kink-anti-kink pair.

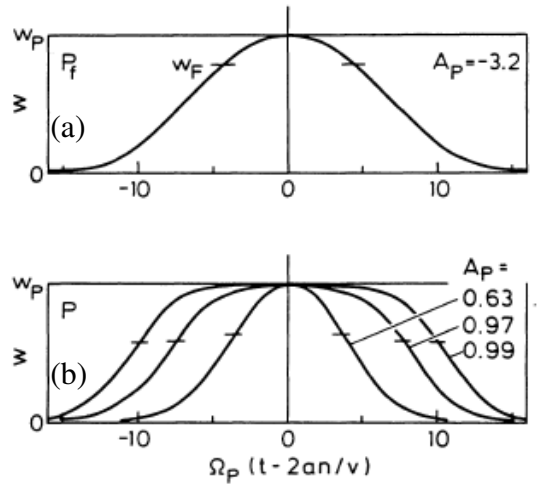

**Figure 8** Fast (a) and slow (b) paraelectric waves. For $A_P \rightarrow 1$ the slow pulse becomes infinitely wide and degenerates into a kink – antikink pair. The marks indicate the amplitude $w_F$ [after Ref. 62].

The variety of solutions outlined above, do not only describe statics and dynamics of Bloch walls in the ferroelectric phase but also travelling pulse waves. A special solution, namely the pulse exciton, carries an integrated dipole moment which increases with decreasing velocity. In the static limit this refers to the case where the shell is far apart from the core but not yet in the ionization limit. In this situation the dipole moment per unit shell charge $p$ is obtained by:

$$p = \frac{v w_F}{2 a \Omega_F} \int_{-\infty}^{+\infty} d\zeta \, \frac{w}{w_F} \tag{53}$$

with $\zeta$ being a dimensionless variable. For small velocities equation 53 reduces to:

$$p \cong (-2 f^2 f' / [8 g_2 g_4 \{2 f' + f\}])^{1/2} (v_2 / v) \tag{54}$$

and yields for $SrTiO_3$ values of the polarization of the order of several nm.

An interesting application can be related to travelling pulse solutions since these can carry large dipole moments and consequently large amounts of energy.

### b) Periodic lattice solutions

Completely different solutions of the nonlinear model can be obtained by considering exact solutions on the periodic lattice [67]. These have been termed periodons and are present on the 3D lattice [68] as well as in the 1D analogue. The solutions are obtained by writing the displacements in the following form ($x = u, v, w$):

$$\vec{x}(L) = \mathrm{Re}\left[\vec{X}_1 \exp\{i(\omega t - \vec{q}\vec{R}(L))\} + \vec{X}_3 \exp\{3i(\omega t - \vec{q}\vec{R}(L))\}\right] \tag{55}$$

where $\vec{X}_1, \vec{X}_3$ are determined by the equations of motion. The dispersion relation for the periodons is given by:

$$\omega_p(\vec{q}) = \tfrac{1}{3}\,\omega_R(3\vec{q}) \tag{56}$$

$\omega_R(q)$ is the dispersion relation in the limit where the polarizability effects are unimportant. The 3D version of the model has been applied to SnTe which is a ferroelectric semiconductor with rocksalt structure and $T_c$=80K. The simple structure is especially well suited to study the periodon dispersion for this system. These are displayed in figure 9 together with the corresponding phonons [69, 70].

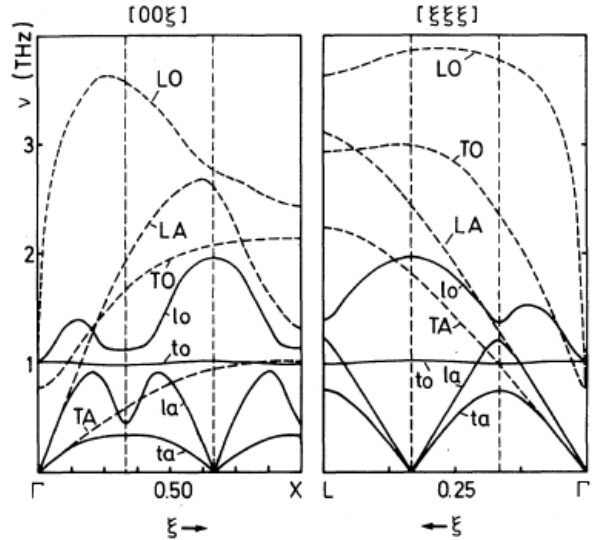

**Figure 9** Dispersion curves of phonons [69, 70] (dashed lines) and periodons (solid lines) in SnTe at T=100K. Capital (small) letters denote the polarization of phonons (periodons).

In the 1D analogue of the model and by restricting this to the case with only the polarizable unit being in the unit cell, the equations of motion reduce to [66]:

$$M\ddot{u}_n = g_2 w_n + g_4 w_n^3 + f'(u_{n+1} + u_{n-1} - 2u_n) \tag{56a}$$

$$0 \quad = -g_2 w_n - g_4 w_n^3 + f(v_{n+1} + v_{n-1} - 2v_n) \tag{56b}$$

The periodon solutions are then given by:

$$\tag{57a}$$

$$\tag{57b}$$

$$w_n = A \sin(\omega t - nqa)$$

$$u_n = B \sin(\omega t - nqa) + C \sin 3(\omega t - nqa)$$

which lead to the periodon dispersion relation:

$$M\omega_p^2(q) = \frac{4}{9}(f + f')\sin^2(3qa/2) \tag{58}$$

The amplitude of the polarizability coordinate is expressed as:

$$A^2(q) = \frac{4}{3g_4}\left[-g_2 - M\omega_f^2(1 - \frac{\omega_f^2}{\omega_R^2 - \omega_p^2})\right] \tag{59}$$

As before, $\omega_R^2$ is the rigid ion limit and $\omega_f^2 = 4f\sin^2 qa/M$. Similar expressions for $B$ and $C$ are obtained [66]. The periodon solutions are temperature independent solutions and as such not suited to describe phase transitions which are related to finite wave vector anomalies, as, e.g., are observed in $K_2SeO_2$ [71]. Besides of this rather exotic compound also the perovskite ferroelectrics show anomalies at finite wave vector in their acoustic mode dispersion as has been emphasized above. It is worth mentioning that the A15 superconducting systems show similar anomalies as seen in $K_2SeO_4$ which have been related to the onset of superconductivity and strong electron-phonon interactions [72 – 74].

In order to include temperature effects in the nonlinear solutions a periodon-phonon coupling scheme has been suggested based on the ansatz [67]:

$$w_n = w_{np} + w_{ns}, \quad u_n = u_{np} + u_{ns} \tag{60}$$

where the subscripts $p$, $s$ refer to the periodon, phonon displacement coordinates. With this superposition principle the equations of motion are modified to:

$$M\ddot{u}_{ns} = (g_T + g_4 w_{np}^2)w_{ns} + f'(u_{ns+1} + u_{ns-1} - 2u_{ns}) \tag{61a}$$

$$0 \quad = (-g_T - g_4 w_{np}^2)w_{ns} + f(v_{ns+1} + v_{ns-1} - 2v_{ns}) \tag{61b}$$

$$M\ddot{u}_{np} = g_T w_{np} + g_4 w_{np}^3 + f'(u_{np+1} + u_{np-1} - 2u_{np}) \tag{61c}$$

$$0 \quad = -g_T w_{np} - g_4 w_{np}^3 + f(v_{np+1} + v_{np-1} - 2v_{np}) \tag{61d}$$

Due to the coupling between phonons and periodons the local electron lattice interaction term in the phonon equations of motion depends on the periodon amplitude while in the periodon equations of motion the amplitude becomes temperature dependent. For the phonons two regimes can be distinguished, namely the low temperature regime which is dominated by static periodons, $\omega_p = 0$ at $q \neq 0$, and the high temperature regime where self-consistent phonons are present in a fluctuating periodon field. In the former regime a site-dependent electron-lattice coupling results:

$$g_T + 3g_4 w_{np}^2(q = 2\pi/3) = \begin{cases} g_T, n \equiv 0 \ (\text{mod}\, 3) \\ -2g_T - 9ff'/(f + f'), n \equiv 1 \ or \ 2 \ (\text{mod}\, 3) \end{cases} \tag{62}$$

which causes a tripling of the lattice constant. For high temperatures the coupling becomes q-dependent and the periodon amplitude $A$ is temperature dependent since the harmonic coupling $g_2$ is replaced by $g_T$:

$$g_T + 3g_4 w_{np}^2(q) = g_T + 3/2g_4 A_T^2(q). \tag{63}$$

This rather simple model accounts well for the temperature dependence of the acoustic mode anomaly observed in $K_2SeO_4$ [71]. The incommensurate wave vector, at which the mode softens, can be obtained by attributing an additional temperature dependence to the core-core coupling $f'$.

Solutions on the lattice as described above will be summarized in the following chapter where, however, in contrast to the above problem, they are no more tied to the lattice periodicity but induce a breakdown of the Bloch theorem accompanied by inhomogeneity and local mode formation [72]. These solutions are believed to be relevant for relaxor ferroelectrics, which are composite materials with locally polarized regions in a soft matrix.

### c) Local modes and discrete breathers

The variety of nonlinear solutions existing in the polarizability model also includes solutions obtained in $\Phi_4$-models where intrinsically localized modes in terms of discrete breathers have been discussed [76 – 79]. These can be important for energy localization and transport in highly nonlinear media [80, 81]. They are not only academically of interest, but are realized in arrays of Josephson junctions [82 – 84], micromechanical oscillators [85], optical wave guides [86, 87], two-dimensional photonic structures [88, 89], antiferromagnetic chains [90], and solid state mixed valence charge transfer compounds [91 – 93]. The problem with these solutions is their stability which is only guaranteed as long as they lie in the gap between acoustic and optic lattice modes [94]. In addition, the hard-core nonlinearity of these double-well models has raised doubts about their existence [94]. Another problem related to the $\Phi_4$-models is the neglect of charge transfer and hybridization effects which are common to many realistic materials. These latter effects are, however, incorporated in the polarizability model and a broader application of the model to complex transition metal oxides can be expected.

The model Hamiltonian as given in chapter 2, equation 14, is used. Instead of considering the SPA solutions, or solutions of the continuum limit or periodic wave solutions, discrete breather solutions are searched for, which require that their frequency is constant within the spatial existence region of the breather and zero outside. The displacement coordinates are assumed to be time periodic, but translational invariance is not required. Correspondingly the displacement coordinates are given by:

$$u_{1n}(t) = A\xi_{1n}\cos(\omega t) \tag{64a}$$

$$u_{2n}(t) = B\xi_{2n}\cos(\omega t) \tag{64b}$$

$$w_{1n}(t) = C\eta_{1n}\cos(\omega t) \tag{64c}$$

Here $A, B, C$ are the frequency dependent amplitudes and $\xi, \eta$ the corresponding displacements. The resulting equations of motion are:

$$-M_1\omega^2\xi_{1n}A = f'A(\xi_{1n+1} + \xi_{1n-1} - 2\xi_{1n}) + g_2C\eta_{1n} + g_4C^3\eta_{1n}^3[\cos(\omega t)]^2 \tag{65a}$$

$$-M_2\omega^2\xi_{2n}B = fC(\eta_{1n+1} + \eta_{1n}) + fA(\xi_{1n+1} + \xi_{1n}) - 2fB\xi_{2n} \tag{65b}$$

$$0 = -g_2C\eta_{1n} - g_4C^3\eta_{1n}^3[\cos(\omega t)]^2 - 2fC\eta_{1n} - 2fA\xi_{1n} + fB(\xi_{2n-1} + \xi_{2n}) \tag{65c}$$

The number of degrees of freedom can be reduced by displacing the equation in $\xi_{2n}B$ by one lattice constant and adding it to the undisplaced one and twice the adiabatic equation:

$$(M_2\omega^2 - 2f)[\{g_2 + g_4\eta_{1n}^2 C^2[\cos(\omega t)]^2\}\eta_{1n}C + C\eta_{1n} + A\xi_{1n} =$$
$$-f^2 C(\eta_{1n+1} + \eta_{1n-1} - 2\eta_{1n}) - f^2 A(\xi_{1n+1} + \xi_{1n-1} - 2\xi_{1n}) \tag{66}$$

The system of equations in $\xi, \eta$ can be solved analytically by using the rotating wave approximation [95, 96] and retaining only the first expansion term. The potential is hard core nonlinear if $g_2 < 0, g_4 > 0$, and soft for $g_2 > 0, g_4 > 0$. Since in the present case the harmonic coupling is attractive and the nonlinear term repulsive, the potential is at this stage hard core nonlinear. The coupled equations of motion to be solved are thus given by:

$$-M_1\omega^2 A\xi_{1n} = f A(\xi_{1n+1} + \xi_{1n-1} - 2\xi_{1n}) + g_2 C\eta_{1n} + \tfrac{3}{4}g_4 C^3\eta_{1n}^3 \tag{67a}$$

$$-M_2\omega^2\{g_2 C\eta_{1n} + \tfrac{3}{4}g_4 C^3\eta_{1n}^3 + 2f(C\eta_{1n} + A\xi_{1n})\} =$$
$$-f^2 C(\eta_{1n+1} + \eta_{1n-1} - 2\eta_{1n}) - f^2 A(\xi_{1n+1} + \xi_{1n-1} - 2\xi_{1n}) + 2f(g_2 C\eta_{1n} + \tfrac{3}{4}g_4 C^3\eta_{1n}^3) \tag{67b}$$

Since the polarizability coordinate $w = v - u$ has been introduced, solutions with $v = u$ are trivial and correspond to the harmonic case. With the ansatz $v = -u$, i.e., $w = -2u$, interesting solutions are obtained which admit a rather transparent analytical analysis. As a first choice for intrinsically localized modes odd parity solutions are considered which correspond to discrete breathers as displayed in figure 10.

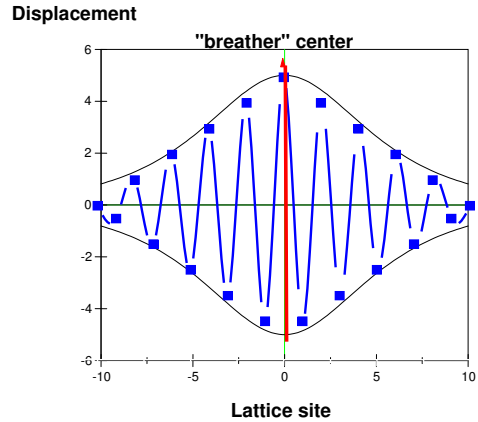

**Figure 10** Displacement of ion 1 with mass $M_1$ within the breather region (blue lines). The breather center is marked by the red line. The black lines correspond to an envelope function proportional to $|1/\cosh|$.

At site $n = 0$ a dipole moment with "length" (displacement $\xi_0$) $n_c/2$ is formed which is compensated by the surrounding lattice by creating at site $n$, counted from the induced dipole moment, a dipole moment in the opposite direction with displacement $\xi_n$ according to: $\xi_n = \xi_0 - n/2 : n \equiv even; \xi_n = -[\xi_0 - n/2] : n \equiv odd$. At lattice site $n_c$ the breather spatial extensions have reached their limit. Thus the spread of the breather critically depends on the magnitude of the dipole moment at the center, the bigger this is the larger is the breather's spatial spread. The frequencies corresponding to these solutions must have the same value within the breather region and are both zero at $n = n_c$. This condition is obeyed over the range $2n_c$ - with small deviations for the $n = 0$ solution - when $g_4$ becomes site dependent like $\tilde{g}_4^{(n)} = g_4 / 2(n - n_c)^2$. The site dependence of the nonlinear coupling constant evidences that in the breather center the potential is steep and approaches the harmonic limit with increasing distance from the center. Also, it is obvious from the breather solution that in the center the electron is far apart from its core, i.e., highly delocalized and extending over the whole breather region, whereas it is tightly bound at the breather's boundaries. The problem is thus related to a spatially extended polaron where the electron travels with a displacement cloud. Opposite to previous breather type solutions [76, 81, 95, 96] this solution is not restricted to nearest neighbor sites only, but covers a rather broad spatial extent which is limited by the associated strain energy. With the definition $g = 2g_2 + 3/4\,\tilde{g}_4 C^2$ the frequencies associated with the displacement pattern at site $n = 0$ are given by:

$$\omega_1^2 = \frac{1}{M_1}\left[4f' + g\,\frac{C}{A}\right] - \frac{2f'}{n_c M_1} \tag{68a}$$

$$\omega_2^2 = \frac{2f}{M_2}\left[1 - \frac{1}{n_c}\frac{1}{2 + \dfrac{gC}{f(2C - A)}}\right] \tag{68b}$$

At all other sites the frequencies are:

$$\omega_1^2 = \frac{1}{M_1}\left[4f' + g\,\frac{C}{A}\right] \tag{69a}$$

$$\omega_2^2 = \frac{2f}{M_2} \tag{69b}$$

In order to confirm the stability of the breather solutions, the frequencies (eqs. 69) have to be compared to those derived from the SPA, eq. 20, since – as outlined above – stable solutions are only realized as long as the solutions lie in the optic-acoustic mode gap. Since $\omega_2$ corresponds to the zone boundary optic mode frequency of the SPA solutions it can be ruled out from further considerations, $\omega_1$ is stable as long as the condition $(2\tilde{f})^{1/2} < [gC/A]^{1/2} < [2\tilde{f}(M_1 + M_2)/M_2 - 4f']^{1/2}$ is fulfilled. This holds when i) the mass ratios of the two sublattices are appropriate, ii) the elasticity $\propto f'$ is soft, iii) the stability of the SPA solution $\propto \tilde{f}$ is guaranteed. From i) it is clear that a light rigid ion mass $M_2$ favors the breather stability. ii) the softer the system the broader is their existence window. iii) real phase transitions in the long wave length limit should not occur.

Since the breather solutions are not independent of the regular lattice – they are imbedded in a lattice matrix – their interaction with the plane wave spectrum has to be taken into account. This can be achieved in an analogous way as has been done in the previous chapter, by using a linear superposition of the breather and regular lattice displacements:

$u = u_{SPA} + u_{breather}$, $w = w_{SPA} + w_{breather}$ with the subscripts referring either to the SPA or breather displacements. This superposition has the effect of modifying the plane wave spectrum locally by the breather solution as is shown in figure 11.

There, three regimes can be differentiated: the plane wave spectrum far away from the breather center (I), the breather center and its spatial spread (II), and, a regime where both interact, corresponding to a charge transfer region (III). Since the solutions are dynamical, the breather can travel through the lattice by riding on the plane waves and induce polar nano-regions which are important in the context of relaxor ferroelectrics.

The displacement superposition principle has the consequence that the embedding matrix is stabilized by the breather since the coupling $g_T \rightarrow g_T + 3g_4 w_{breather}^2$, whereby a polarizabilty catastrophe $g_T = 0$ can be inhibited. On the other hand the breather adopts a temperature dependence since $g_2$ is replaced by $g_T$. The coupled system exists now of the two SPA modes and the breather mode, which – of course – is spatially limited and appears not as superlattice reflections but only as enhanced diffuse scattering or ghost mode.

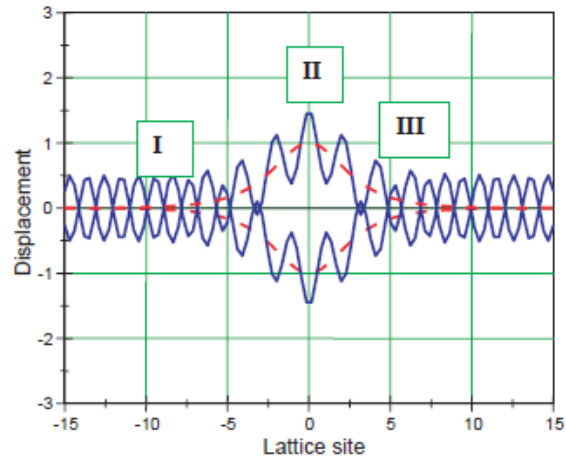

**Figure 11** Superposition of the regular lattice (I) and breather displacements (II). Region III is the regime where interactions between I and II are relevant. The red dashed line corresponds to an averaged envelope function.

The interplay between the breather solution and the regular lattice has been studied numerically by first solving for the coupled equations of motion including the temperature dependence of the involved modes. Typical dispersion curves are shown in figures 12, 13. Here the temperature at which the DB emerges from the continuum of the optic mode spectrum is marked as $T_{onset}$, and a second temperature scale is set by the temperature $T_m$ where the DB starts to interact with the acoustic mode. At this latter temperature huge effects on the elastic constants set in since the acoustic mode merges into the DB mode at finite momentum values which are temperature dependent due to the lower energy of the DB as compared to the acoustic mode. Specifically, a collapse of the elastic constants is expected as soon as a certain critical q-value is reached. Such a behaviour, strongly reminiscent of ferroelastic materials, has been observed in relaxor ferroelectrics, however never marking a real ferroelastic transition, but rather a diffuse one [97]. This can be attributed here to the fact that the mode-mode coupling q-value is temperature dependent and smoothly approaches a critical one.

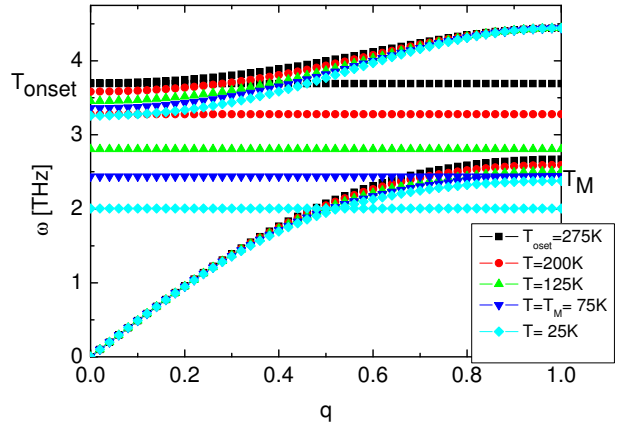

**Figure 12**: Transverse optic, acoustic, and breather mode frequencies as functions of wave vector $q$ and for various temperatures as indicated in the figure. The above case refers to a dense breather system. In this case $f \gg g_T$.

$T_{onset}$ as well as $T_m$ depend on the choice of parameters and both may extend to much higher as well as much lower temperatures than shown in figures 12 and 13. Note that with decreasing temperature the optic zone center and the acoustic zone boundary modes soften slightly but both remain stable for all temperatures, whereby a ferroelectric type transition is prohibited. Also this behavior has been observed in relaxor ferroelectrics [98], [99]. In figure 12 $f$ is much larger than $g_T$, which leads to the positive dispersion of the optic mode with increasing q-value and the merging of the DB with the optic mode phonon continuum at large wave lengths. In the opposite case, i.e. $g_T \gg f$, the optic mode dispersion is reversed and the breather formation is first observed at the zone boundary, where it splits off from the plane wave states at small temperatures to decrease in energy with increasing temperature (figure 13). In this case $T_m$ is larger than $T_{onset}$.

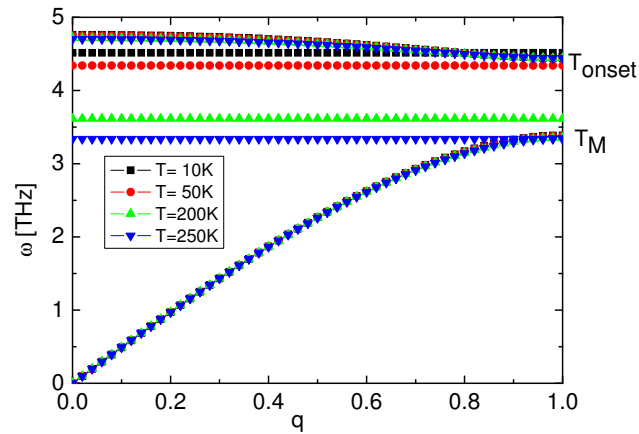

**Figure 13** Typical dispersion curves for the coupled system (SPA plus DB) as functions of temperature. In this case $f << g_T$ and $\omega_{TO}^2(q=0) > \omega_{TO}^2(q=2\pi/a)$ and the reversed temperature dependence of the breather mode as compared to fig. 12 is observed, i.e. the breather mode energy decreases with increasing temperature.

The consequences of breather formation for ferroelectrics will be outlined in the following chapter where their role is discussed in the context of relaxor ferroelectrics. An important issue of breathers is that they are not tied to the lattice periodicity but explicitly break the translational invariance and enrich the physics of ferroelectrics.

## 5. Relaxor ferroelectrics

Relaxor ferroelectricity (RF) is almost exclusively observed in doped $ABO_3$ systems where either the A or B site or both are substituted by another cation or transition metal element. Typically, the dielectric constant becomes then strongly frequency dependent and peaks at a frequency dependent temperature. As a consequence they either do not show any phase transition or a broad distribution in $T_C$. Characteristic for these compounds is that deviations from linear in T dependencies occur at temperatures much higher than the dielectric peak, which has been named the Burns temperature [100]. Already in 1960 Smolenski [101] proposed that relaxor ferroelectricity is caused by chemical inhomogeneity and thus gives rise to diffuse phase transitions (DPT) with a "spatial distribution of first order phase transition temperatures". However, also non RF's can exhibit DPT's [102] which questions the approach of Ref. 101. Cross [99] suggested that nanoscale thermally fluctuating dipoles in polar nano regions (PNR) are the origin of RF, motivated by the discovery of nanoscale Mg/Nb chemical ordering in PMN [103]. Since RF behavior persists when the chemical ordering occurs over a macroscopic scale [104], this suggestion is incomplete. The analogy of RF's with polar glasses [105] led to the introduction of the random field Ising model [106] which was subsequently extended by Blinc et al. into the spherical random bond random field model [107]. The essential ingredients of both models are based on the assumption that random interactions between the PNR's take place which successfully reproduced the radio-frequency response and NMR results [108]. Both approaches are phenomenological and do not provide any microscopic insight into the physics.

To gain insight into the latter the intrinsic strong nonlinearity of the systems has to be taken into account [37, 39, 109, 110] which is the source of the formation of intrinsic local modes (ILM) in terms of discrete breathers [76 – 79]. The model applies to dilute ferroelectrics as, e.g., Ca doped $SrTiO_3$ [13, 111] in which ILM's represent Ti ions in the vicinity of Ca, whereas the remaining lattice corresponds to the embedding matrix. Other candidates are for instance PMN, PZT, and related compounds, where the ILM refers to the Pb polarization in the oxygen ion cage. The strong hybridization of the Pb p- and oxygen ion p-orbitals induces a shortening of their bond length ($\approx 2.4$Å) as compared to the average distance of 2.8Å. Thus Pb forms a short bond with 3-4 oxygen ions and becomes off-centered in the oxygen ion cage by as much as 0.5Å [112] whereas the $Mg/NbO_3$ lattice corresponds to the soft dielectric matrix. The random occupation of the lattice sites by $Mg^{2+}$ and $Nb^{5+}$ hinders the direct coupling of the ILM's and reduces the susceptibility.

By applying the previous approach of Chapter 3c for the superposition of breather and plane wave displacements, the temperature dependencies of the involved modes (transverse optic q=0 mode, transverse acoustic zone boundary mode and breather mode, respectively) have been calculated for two distinctly different cases: i) dense breather distribution corresponding to a large spatial spread of the ILM, and ii) dilute breather distribution with extension limited to a few lattice constants.

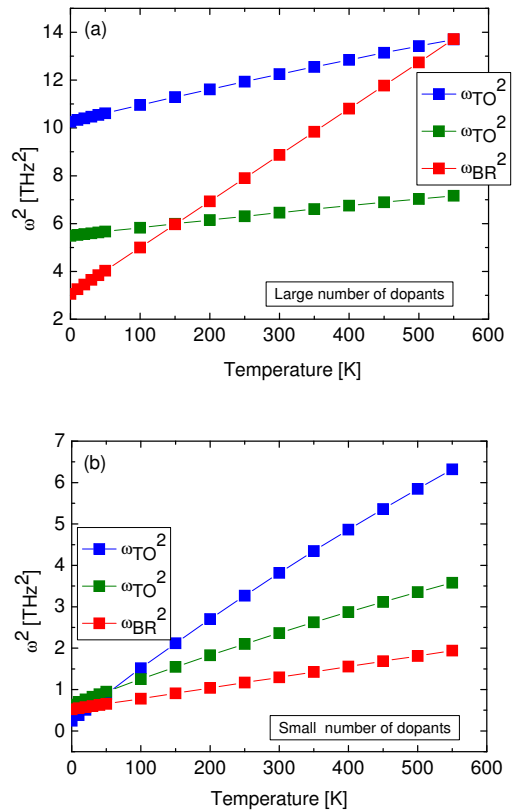

**Figure 14** Squared TO (q=0), TA (q=π/2a) and breather frequencies as functions of temperature. Figure 14 (a) refers to the *dense* case, (b) to the *dilute* case.

In the case of i) dense ILM distribution (figure 14 (a)), the long wave length optic mode as well as the zone boundary acoustic mode show only a minor softening with decreasing temperature, whereas the squared ILM mode frequency is nearly linearly dependent on temperature. This behaviour governs the dynamical properties of the composite system and a structural instability is prevented. Note that the ILM emerges from the optic mode at a temperature, which can be identified with the Burns temperature [100], where the dynamical properties of the RF are dominated by the polar clusters.

Experimental evidence for a realisation of this scenario has been obtained recently by pulsed neutron scattering measurements. Through an energy-integrated powder atomic pair-density function (PDF) measurement, local deviations from cubic symmetry into polar rhombohedral symmetry were found to occur below the Burns temperature. By an energy-resolved dynamic PDF measurement through inelastic pulsed neutron scattering, it was found that the local dynamic Pb polarization of about 0.3Å against the Mg/Nb lattice occurs below the Burns temperature in the energy range of 2-5 THz – in excellent agreement with the results in figures 14 – while the radio-frequency local polarization exists only below room temperature [113]. This scenario is also supported by broad-band dielectric measurements and Fourier transform infrared transmission spectroscopy in an extended frequency range [114].

In the case ii), dilute ILM (figure 14 (b)), it is observed that the squared TO mode frequency softens substantially with decreasing temperature, reminiscent of classical displacive soft mode behaviour, however being nonlinearly dependent on temperature and becoming pinned to the ILM at low temperatures and small energies. Here a first order transition to a polar state is possible and a crossover between RF and displacive behaviour takes place [38]. Opposite to case i), the ILM has now lost its pronounced temperature dependence but contributes to that of the TA zone boundary frequency, which is strongly dependent on temperature due to the mode crossing/coupling of the ILM and the acoustic branch. It is important to note that the ILM does not split off from the optic mode spectrum, but exists at all temperatures and gains largest spectral weight at small momenta, whereas in case i) the spectral weight is transferred to it from the zone boundary TO mode.

The two limiting cases of ILM formation in RF's are realized in various systems where doping is possible over a large range [114 – 116]. While highly doped RF's mostly do not show any structural instability, consistent with the above results, for small doping concentrations a polar instability is frequently observed and even crossovers from RF to displacive behaviour have been reported [114, 117] and discussed in detail recently [38]. Besides the dynamical properties of RF's, electric field induced phenomena and the distribution in relaxation times are also consequences of this approach. By identifying the ILM displacement with a local fluctuating polarization field $P$, field effects can be incorporated through:

$$E = g_2 P + g_4 P^3 + \tau g_2 \frac{dP}{dt}, \tag{70}$$

where $\tau$ is given by a Gaussian distribution as a consequence of the local variations in the potential wells. The results for the dielectric response of the composite structure are shown in figure 15. In both cases a strong frequency and temperature dependence of the dielectric responses are observed, as has also been found experimentally. However, in the dilute limit (figure 15 (a)) the response is sharp and at small energies with a less pronounced frequency dependence. In the dense case (figure 15 (b)) the response is broadened at all temperatures, indicating that the properties of this system are governed by the extended dopant regions.

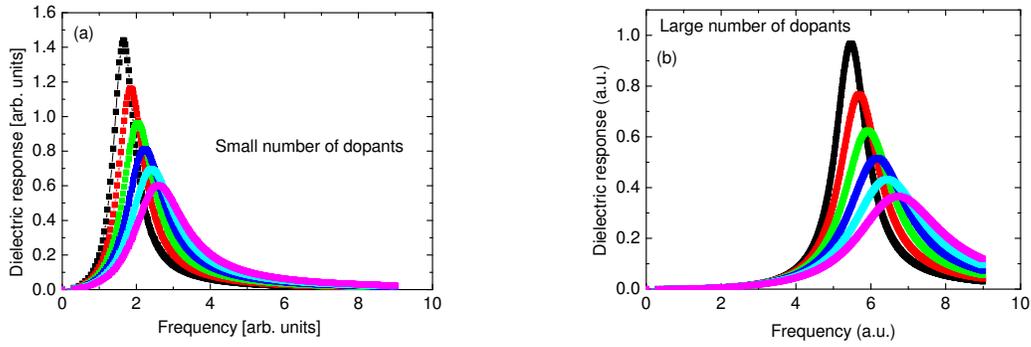

**Figure 15** Dielectric response as a function of energy for various temperature. With decreasing temperature (starting from magenta to black) the peak maximum shifts to smaller energies. (a) refers to the dilute limit, whereas (b) corresponds to the dense case.

Below $T_C$ the potential defining parameters have to be replaced by $-2g_2, -2g_4$ and $g_T \rightarrow -2g_T$ [17]. This replacement yields the typical recovery of the soft mode below $T_C$ which hardens with decreasing temperature. In the presence of the ILM this dependence is reduced since now – analogous to the paraelectric phase: $2g_2 \rightarrow 2g_2 + 6g_4C^2$. To make things explicit, the temperature dependence of $\omega_f^2$ in the ferroelectric phase is given by:

$$\omega_f^2 (T \leq T_c) = \frac{2f[2g_2 - 3g_4(2 <w^2>_T - C^2)]}{\mu[2f + 2g_2 - 3g_4(2 <w^2>_T - C^2)]} \quad (71)$$

From Eq. 71 it becomes clear that the soft mode hardens below $T_c$ but its temperature dependence can be quite different from the case of a displacive ferroelectric caused by the temperature reducing effect from the ILM amplitude. A typical T dependence of $\omega_f^2 (T)$ is shown in figure 16 where with increasing ILM amplitude the temperature dependence of $\omega_f^2$ is systematically reduced. The Curie constant is substantially suppressed with increasing ILM amplitude as seen experimentally [118 - 121]. Below $T_c$ the soft mode hardens and this hardening gets less pronounced with increasing breather amplitude.

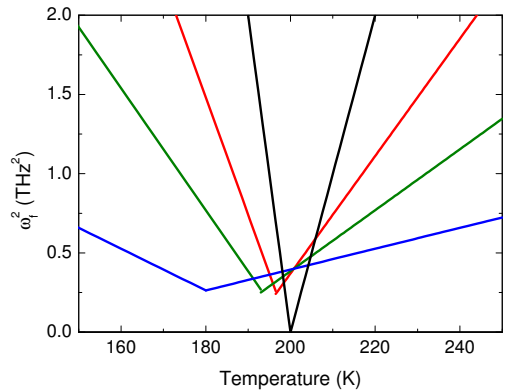

**Fig. 16** Temperature dependence of the squared soft mode frequency without ILM renormalization (black) $T_c$=200K, small ILM amplitude (red), large ILM amplitude (green), and the extremely dense ILM case (blue).

The two limiting cases of dense and dilute ILM's can adopt all intermediate possibilities since the ILM amplitude varies continuously with doping. This means, that a smooth crossover from the purely displacive to the relaxor case is possible, as also suggested from experiments [113, 118]. In the extreme dilute case, considered above, even a coexistence of relaxor and soft mode dynamics can occur. The observation of inhomogeneous polar nanodomains, discussed here in terms of local mode formation, is not tied to $T_B$ but is expected to be possible above $T_B$. At these high temperatures, coherence, which causes ILM formation, of the polar nanoregions is not yet achieved making it difficult to observe them. They should, however, be observable through local probe techniques such as NMR, PDF or EXAFS. The dynamics of the matrix are expected to be "normal". Only, when the breather frequency reaches the phonon spectrum at either the zero momentum optic phonon or the transverse acoustic zone boundary mode, namely at $T_B$, coherence takes place and the polar nanoregions are observable by conventional scattering techniques.

A direct probing of the continuous crossover from strong mode softening in ferroelectrics to an incomplete softening relaxor is still lacking., even though the incomplete mode softening has been observed by infrared and broadband dielectric spectroscopy in various relaxor systems [114, 122, 123, 124]. Indeed, observing the continuous crossover, would require a full set of single crystals in a solid solution (for example, the $BaTi_{1-x}Zr_xO_3$ (BTZ)) solution. Reliable IR, RAMAN and inelastic neutron scattering would then be available. Since, however, such high quality single crystals are still missing, TO phonons investigations in ceramics samples could not evidence the required slowing down [121]. An alternative way can be found in the high temperature extrapolation of the dielectric permittivity in a set of ceramics like BTZ. Indeed, through the Lyddane-Sachs-Teller relation, the low frequency permittivity is directly related to the TO mode softening: $\varepsilon(T) \propto (\omega_f(T))^{-2}$ which is valid for $T > T_B$ in relaxor ferroelectrics. Instead of focusing the attention to the critical temperature range where either true ferroelectric or dispersed relaxor behavior occurs, we investigated the temperature range where the Curie-Weiss law is valid in both instances:

$$\varepsilon(T) = \frac{C_{CW}}{(T - T_c)} \tag{72}$$

with $C_{CW}$ being the Curie constant and $T_C$ is an extrapolated temperature which loses its meaning in the case of relaxors because it is first frequency dependant and second the deviation from equation 72 precludes any reliable fitting of the permittivity data.

Usually, it is this deviation which is used to assign the occurrence of a relaxor state. On the basis of the breather model temperatures where equation 72 is valid for ferroelectrics as well as for relaxor are of interets, i.e. for the latter systems above $T_B$. Plotting the inverse of $\varepsilon(T)$ versus T is the correct way to evidence the Curie-Weiss law. This is done on Fig. 16 for some key compositions in the BTZ family. As the critical temperature range has been omitted in these plots, only the computed slope C of the lines is the parameter of interest. This Curie constant is reported in Table 2 for several BTZ compounds together with their ferroelectric or relaxor characteristic transition temperatures [118]. In the same table typical Curie constants for $KNbO_3$ and $PbTiO_3$ ferroelectrics are also included. Obviously, the Curie constants split into two groups: the one of displacive type ferroelectrics with $C_{CW} > 1.5.10^5$K and the one of relaxors with $C_{CW} < 1.1 \ 10^5$K. Even if this difference is not huge, it is systematically observed in the $BaTiO_3$ based compounds including the BaCaTiZrO (BCTZ) ferroelectric/relaxor compositions (table 2). This is an indirect confirmation of the key model expectation of figure 16 calling for an incomplete softening of $\omega_f^2$ in relaxors. The clue here is that the variation of Curie constants is quite sharp when shifting from the ferroelectric to relaxor compositions while the critical temperatures are continuously decreasing without any sign of criticality for many lead free solid solutions [118]. This a very strong support for the breather model – whether diluted or dense – being a microscopic precursor of two different behaviors which are only macroscopically distinguishable close to the critical temperature. The agreement between these experimental findings and the model predictions becomes clear by comparing figures 17 (a) and 17 (b).

While in the above the possibility of a smooth crossover from displacive to relaxor behavior has been demonstrated to exist, in the following anomalous features observed in the dielectric response of relaxors are addressed [39]. These are directly related to the ILM formation which introduces new time and length scales as compared to the average structure. The spatial extent of the ILM's can be related directly to PNR's and is shown to cause the enhanced diffuse scattering as observed by inelastic and pulsed inelastic neutron scattering [113, 126 – 128], Raman spectroscopy [129], X-ray scattering [113], and NMR measurements [130].

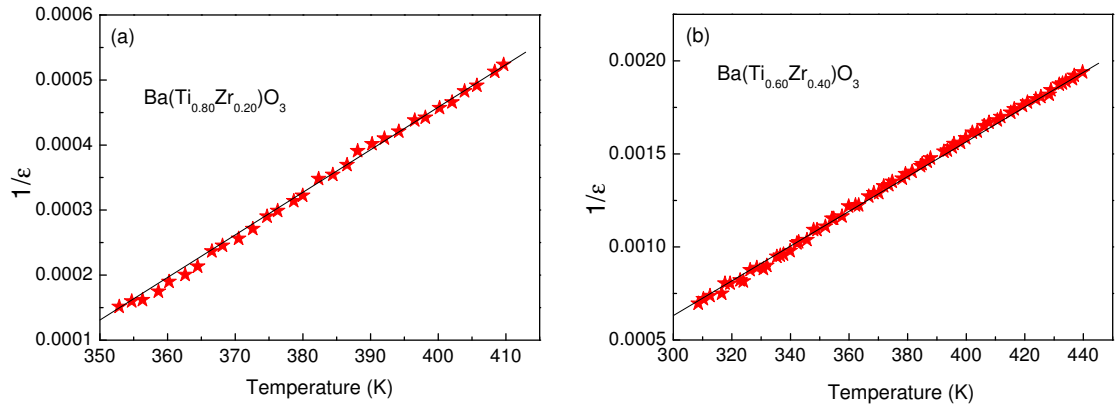

**Figure 17** Curie Weiss behavior of (a) a ferroelectric $Ba(Ti_{0.80}Zr_{0.20})O_3$ and (b) a relaxor $Ba(Ti_{0.60}Zr_{0.40})O_3$ taken from the whole family of materials reported in Tab. 1. The lines are obtained through a fit to Eq. (72). The data have been taken at 1 THz.

| Composition | $T_c$ or $T_m$ (K) | Curie Constant $C_{CW}$ (K) | Ref. |
|---|---|---|---|
| $BaTiO_3$ **(F)** | 400 | 1.5E5 | [125] |
| $Ba(Ti_{0.85}Zr_{0.15})O_3$ **(F)** | 340 | 1.5E5 | [120] |
| Ba(Ti$_{0.80}$Zr$_{0.20}$)O$_3$ (**F**) | 314 | 1.6E5 | [120] |
| | | | |
| Ba(Ti$_{0.63}$Zr$_{0.37}$)O$_3$ (**R**) | 194 | 1.06E5 | [120] |
| Ba(Ti$_{0.60}$Zr$_{0.40}$)O$_3$ (**R**) | 188 | 1.03E5 | [120] |
| Ba$_{0.88}$Ca$_{0.12}$(Ti$_{0.63}$Zr$_{0.37}$)O$_3$ (**R**) | 169 | 9.9E4 | [120] |
| Ba$_{0.90}$Ca$_{0.10}$(Ti$_{0.70}$Zr$_{0.30}$)O$_3$ (**R**) | 209 | 8.2E4 | [120] |
| | | | |
| KNbO$_3$ (**F**) | 691 | 2.4E5 | [125] |
| PbTiO$_3$ (**F**) | 763 | 4.1E5 | [125] |

**Table 2**: Curie constants $C_{CW}$ of several ferroelectric (F) and relaxor (R) materials in the same BaTiO$_3$ family; some other ferroelectric materials are listed for reference. T$_c$, T$_m$ are the real, respectively, extrapolated transition temperatures.

In addition, infrared and broad band spectroscopy experiments [122, 123, 131] report the observation of additional structure which is incompatible with the crystal symmetry. In order to understand these latter experiments the complex dielectric function of the coupled system is calculated within standard theory:

$$\tilde{\varepsilon}(\omega) = \varepsilon_\infty + \sum_i S_i R_i(q, \omega), \qquad (73)$$

where the $\omega_i$ (i=1, 2) refer to the momentum $q$ dependent optic lattice mode and the DB mode frequencies, and $R_i(q, \omega)$ is the response function for the corresponding mode. In anharmonic crystals this is expressed as:

$$R_i(q, \omega) = [\omega_i^2(q) - \omega^2 + 2\omega_i(q)\sum_i(q, \omega)]^{-1}, \qquad (74)$$

with $\sum_i(q, \omega)$ being the self-energy. In the above case this energy relates directly to the integrated dipole moment which is proportional to the self-consistently derived thermal average $\left\langle w^2 \right\rangle_T = \sum_{q,i} \hbar w^2(q,i)/[M_i\omega_i(q)]\coth\dfrac{\hbar\omega_i(q)}{2kT}$. Since the DB mode frequency is dispersionless by definition, the response function is calculated in the limit $q \to 0$. The calculated imaginary part of the dielectric constant is shown for small DB spatial extents and various temperatures in figure 18.

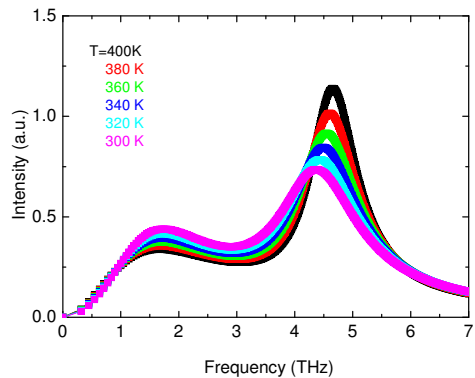

**Figure 18** Temperature and frequency dependence of the breather mode response for fixed breather extent. The choice of the extent corresponds to the dilute doping limit.

While at high temperatures the dielectric response is almost completely dominated by the lattice mode pseudo-harmonic frequency, with decreasing temperature an additional peak develops on the low frequency side which gains intensity on expense of the lattice mode. Together with the intensity redistribution an anomalous broadening of the lattice response sets in whereas the DB mode sharpens. The appearance of the DB in the low frequency regime suggests a strong DB acoustic mode coupling leading to enormous anomalies in the elastic constants as observed experimentally [132 − 138]. Upon increasing the DB spatial extent, all the above described features become much more pronounced (figure 19) and the DB mode dynamics slow down substantially. This has the consequence that the larger the PNR's are, the more they approach an almost static limit where two time and length scales appear. Also, a much stronger piezoelectric coupling is present in the latter case which could be the cause of the enhanced piezo response reported for relaxor ferroelectrics [139, 140].

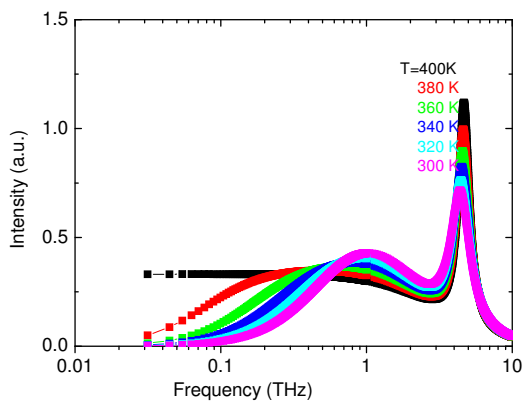

**Fig. 19** Temperature and frequency dependence of the breather mode response for fixed breather extent. The choice of the extent corresponds to the dense doping limit.

Since the DB stabilizes the lattice mode frequency a soft mode induced transition is suppressed, but only a minor (still feasible) temperature dependence of the lattice mode remains which is independent of the DB extent (figure 14 (a)). Opposite to this finding, the DB mode frequency depends strongly on the DB spatial extent and softens when the extent is large and approaches the lattice mode spectrum for the opposite limit (figure 14 (b)).

The theoretical picture can be directly related to recent extended data on the complex permittivity obtained for 0.2PSN-0.4PMN-0.4PZN ceramics which are detailed below. These experiments have the advantage to cover a large frequency regime whereby coexistence regions of slow and fast dynamics can be detected.

In order to understand the dielectric relaxation in relaxors, it is more convenient to use frequency plots of the complex permittivity at various representative temperatures (figure 20). Obviously a huge change in the dielectric dispersion takes place with decreasing temperature. At temperatures $T \geq 400$ K, the dielectric loss dispersion is clearly symmetric and observable only at frequencies larger than 1 GHz. On cooling, the relaxation slows down and broadens. At temperatures around 300 K the relaxation becomes strongly asymmetric and very broad. On further cooling, the dielectric dispersion becomes so broad that only part of it is visible in the available frequency range.

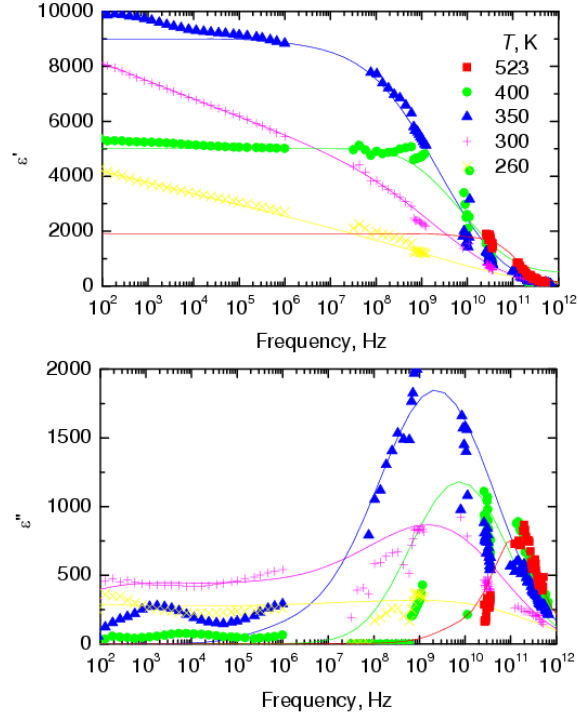

**Figure 20** Frequency dependence of complex dielectric permittivity of 0.2PSN-0.4PMN-0.4PZN ceramics measured at different temperatures. The solid lines are the best fits with the distribution of the relaxation times.

In order to determine the real and continuous distribution function of relaxation times $f(\tau)$ the Fredholm [141] integral equation is used:

$$\varepsilon^{*}(\nu) = \varepsilon_{\infty} + \Delta\varepsilon \int_{-\infty}^{\infty} \frac{f(\tau) d \log \tau}{1 + 2\pi i \nu \tau} \tag{75}$$

Where the Tikhonov regularization method has been employed [142 – 144]. This method and calculation technique is described in detail elsewhere [145].

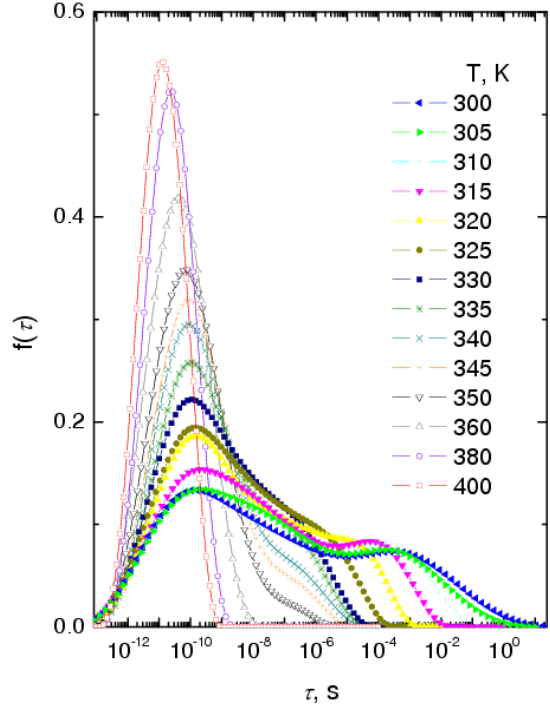

**Figure 21** Distribution of relaxation times f (τ) of 0.2PSN-0.4PMN-0.4PZN calculated at different temperatures

The calculated distributions of the relaxation times $f(\tau)$ are presented in figure 21. Symmetric and narrow distributions of relaxation times are observed at temperatures T>350K. On cooling $f(\tau)$ adopts an asymmetric shape and a second maximum appears at lower frequencies. The shortest and longest limits of $f(\tau)$ were calculated (level 0.1 of the maximum $f(\tau)$ was chosen for definition of the limits) at various temperatures (figure 21). The maximum relaxation time $\tau_{max}$ diverges according to the Vogel-Fulcher law:

$$\tau = \tau_0 e^{\frac{E_f}{k_B(T-T_0)}} \tag{76}$$

where $\tau_0 = 10^{-12} s$, $T_0 = 285$ K, $E_A / k_B = 715K$. All shortest relaxation times have the same values, namely $10^{-12} - 10^{-13}$ s, and are almost temperature independent within the accuracy of this analysis. The most probable relaxation time $\tau_{mp}$ obtained from the peak of distributions follows the Arrhenius law (figure 22)

$$\tau = \tau_0 e^{-\frac{E_A}{k_B T}} \tag{77}$$

with parameters $\tau_0 = 1.06 * 10^{-15} s$, $E_A / k_B = 3825K$.

The distribution function $f(\tau)$ was determined only at rather high temperatures, when the relaxation time lies within the experimental frequency range, whereas for $\tau_{\max}$ being below this range, neither the static permittivity nor $\tau_{\max}$ can be determined unambiguously. It can, however, be expected that $\tau_{max}$ does not diverge below $T_0$, since some dispersion also remains below the freezing temperature. On the other hand, the Arrhenius law describes well enough the temperature dependence of $\tau_{mp}$ in the considerd temperature range of 300 to 400K. At higher temperatures (T>400) this law breaks down since $\tau_{\max}$ is smaller than $10^{-13}$ s, and stays almost constant for this T regime..

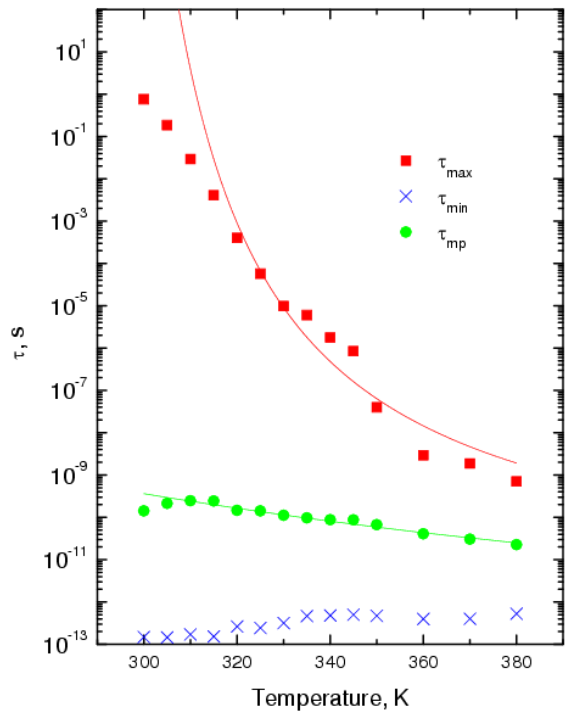

**Fig. 22** Temperature dependences of the longest $\tau_{\max}$, most probable $\tau_{mp}$, and the shortest $\tau_{min}$ relaxation times in 0.2PSN-0.4PMN-0.4PZN ceramics.

Even though the origin of the two-component dielectric relaxation is still under debate, it is interpreted here as arising from the DB formation and its interaction with the embedding matrix. The emergence of the low frequency broad peak is identified with the DB mode frequency which gains strength on expense of the high frequency response with decreasing temperature in close analogy to the modeling (figures 18, 19, and 21). Note, that the high frequency peak is not completely temperature independent but that a slight softening with decreasing temperature sets in as predicted theoretically. The overall qualitative agreement between experiment and theory thus confirms the picture developed in the beginning of this paragraph.

### 6. Isotope effects on the transition temperature

Isotope effects on the ferroelectric transition temperature are experimentally well investigated and also the replacement of any of the involved sublattice masses by isovalent ions is known to affect $T_C$ [16]. However, substantial isotope effects have only been observed in hydrogen-bonded ferroelectrics upon replacement of hydrogen by deuterium [16]. While it was long believed that this effect stems from the proton dynamics only [146 – 148], it was shown more recently to be a consequence of the precise bond geometry [149] and the coexistence of order / disorder and displacive dynamics [150]. In perovskite ferroelectrics isotope effects are absent with the only exception of $SrTiO_3$ where the exchange of $^{16}O$ by $^{18}O$ induces ferroelectricity [151] in the otherwise quantum paraelectric [50 – 52]. This observation has intensified the research on $SrTiO_3$ enormously in order to find a microscopic explanation of it. However, before this discovery it was already shown [33] that isotope effects in perovskites can occur but depend critically on the transition temperature. While for high values of $T_C$ isotope effects are absent, low values of it support them and even increase the isotope effect when $T_C$ tends to zero.

Within the polarizability model the explicit expression for $T_C$ is obtained from the condition that $\omega_F^2 = 0$. This is fulfilled when $g_T = 0$ or equivalently:

$$g_2 + 3g_4 \sum_{q,j} \frac{\hbar}{N\omega_{qj}} w_1^2(qj) \coth \frac{\hbar\omega_{qj}}{2k_B T_C} = 0 \qquad (78)$$

with the core-shell eigenvector being given by equation 27. From equation 78 it is obvious that any mass dependence of $T_C$ vanishes in the limit that $\coth x \approx 1/x$ which holds for large values of the transition temperature. For small values of $T_C$ and replacing the sum in equation 87 by its integral, the mass dependence can be calculated numerically as has been done in [33]. By concentrating on a single parameter set and using the self-consistently derived double-well defining quantities, the change in $T_C$ with the changes in the sublattice masses $M_1, M_2$ is obtained (figures 23 (a), (b)).

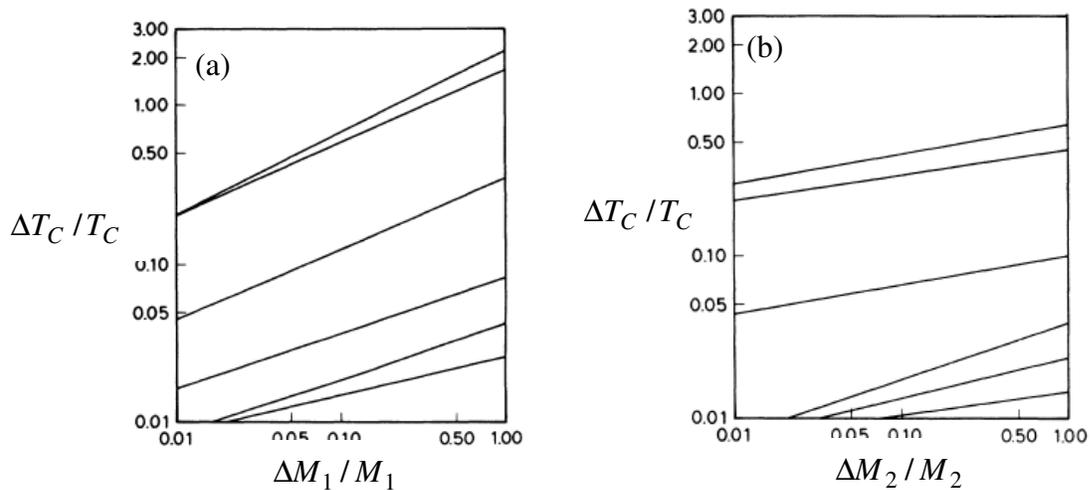

**Figure 23** Double logarithmic plot of $\Delta T_C / T_C$ as a function of $\Delta M_1 / M_1$ (a) and $\Delta M_2 / M_2$ (b) for various values of $|\,g_2\,|$ [after Re. 33].

The variations in $T_C$ have been achieved by changing the depth of the double-well potential, i.e., $|\,g_2\,|$. For a given value of $|\,g_2\,|$ the corresponding transition temperature has been derived, and then $M_1, M_2$ being increased and $T_C$ again calculated. As is obvious from figure 22, $T_C$ always increases with increasing either of the two masses, with the effect being stronger for changes in $M_1$ (the polarizable mass) as compared to $M_2$ (the rigid ion mass). By defining in analogy to superconductivity a critical exponent $\gamma$ like $k_B T_C \approx M_i^{\gamma_i}, i = 1,2$, $\gamma$ is always positive, strongly dependent on $T_C$ and also on $M_i$. For large enough $k_B T_C$ $\gamma$ tends to zero. The explicit dependence of $\gamma$ on $k_B T_C$ results in $\gamma_1 \cong 56 T_C^{-1.43}, \gamma_2 \cong 9.53 T_C^{-1.15}$.

After the discovery of Itoh et al. [151] of isotope induced ferroelectricty in SrTi$^{18}$O$_3$ the above results have been applied to this compound and compared to KTaO$_3$ in order to see whether there is also the possibility to find a finite $T_C$ upon oxygen isotope replacement [152 – 154]. The results for SrTi$^{16/18}$O$_3$ are shown in figure 24, where also the 50% replaced compound is shown.

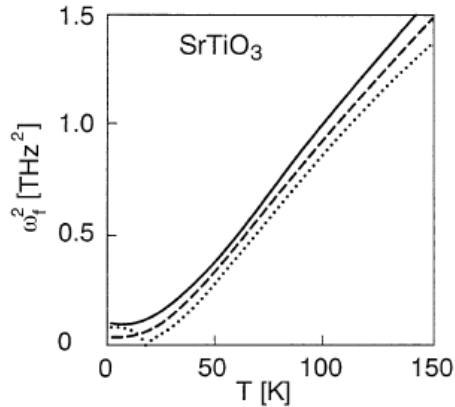

**Figure 24** Temperature dependence of the squared soft mode frequency of SrTiO$_3$. The full line corresponds to the $^{16}$O system, the dotted line to the $^{18}$O system and the 50% substituted compound is represented by the dashed line.

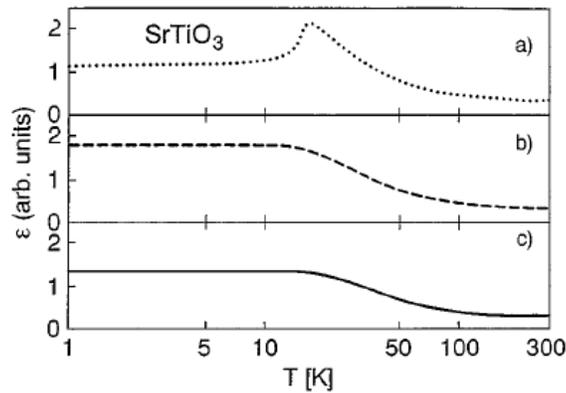

**Figure 25** Static dielectric constant $\varepsilon$ as a function of temperature. The full line corresponds to the $^{16}O$ system, the dotted line to the $^{18}O$ system and the 50% substituted compound is represented by the dashed line.

The corresponding dielectric constants $\varepsilon$ are shown in figure 25. While the $^{16}O$ and the 50% exchanged compounds exhibit a broad plateau below 25K, a peak in the dielectric constant appears for the fully isotopically substituted one at the same temperature as the ferroelectric mode is zero.

The quantum paraelectric behavior of $KTaO_3$ is suggestive to assume that also here isotope induced ferroelectricity could occur. Also the similarity in their local double-well potentials (figure 6) supports tentatively this possibility. However, the equivalent calculation for $KTaO_3$ as carried through for $SrTiO_3$ reveals that $KTaO_3$ remains a quantum paraelectric (figure 26) even if fully oxygen isotope exchanged. This observation can be explained by the fact that polarizability effects are less pronounced in this system as compared to $SrTiO_3$ which is evidenced by the shallower and broader double-well potential (figure 6).

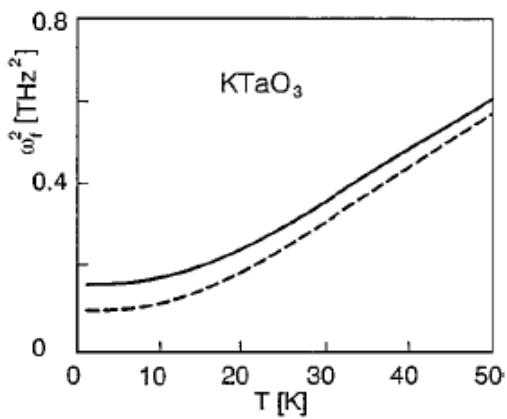

**Figure 26** Temperature dependence of the squared soft mode frequency of $KTaO_3$. The dashed line corresponds to $KTa^{18}O_3$, the full line to $KTa^{16}O_3$.

After Itoh et al.'s discovery [151] the phase diagram of $SrTi(^{16}O_{1-x}{}^{18}O_x)_3$ has been investigated experimentally [155] and theoretically [154]. This was motivated by the fact that controversial results for the dynamics of the end members (x=0., x=1.) have been obtained from different experimental techniques. While long –wavelength probes such as Raman and infrared studies provided evidence for a purely displacive transition with perfect mode softening [155 – 157], local probes as NMR and EPR [158] support an order / disorder driven phase transition. The investigation of the phase diagram of $SrTi(^{16}O_{1-x}{}^{18}O_x)_3$ offers thus a unique opportunity to study the crossover from quantum paraelectric to ferroelectric and possible coexistence of order / disorder and displacive dynamics. The quantum limit of phase transitions has been extensively studied by various approaches. Oppermann and Thomas [159] studied the critical behavior for a $\Phi_4$ model and predicted a change in the scaling relations in the quantum regime. Schneider et al. [160] investigated an n-component vector model and observed that the leading exponent of the dielectric susceptibility changes from 1 to 2. The renormalization group study by Schmeltzer [36] emphasized the enhanced dimensionality of the quantum state as compared to the ferroelectric state. Scaling properties have been investigated for the quantum paraelectric solid solutions of Ca doped SrTiO$_3$ [161] where both dynamical limits, namely order / disorder and displacive have been taken into account. A very different study has been carried through by Rubtsov and Janssen [162] who studied the $\Phi_4$ model in two and three dimensions. From this study a continuous change from soft mode to transverse Ising behavior was found.

The x-dependent phase diagram of SrTi($^{16}$O$_{1-x}$$^{18}$O$_x$)$_3$ has been calculated within the above outlined formalism and $M_1$ was changed linearly with x whereas the remaining model parameters are unchanged (figure 27).

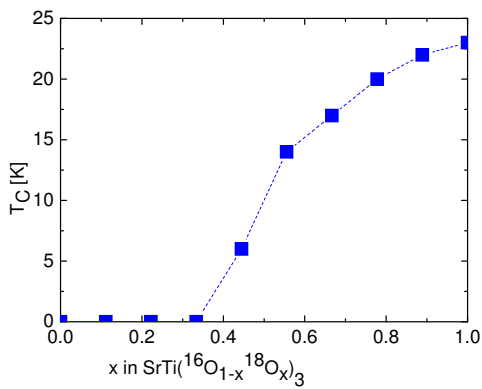

**Figure 27** x-dependence of the ferroelectric transition temperature T$_C$ in SrTi($^{16}$O$_{1-x}$$^{18}$O$_x$)$_3$.

It is important to note that the points in figure 25 where T$_C$ is seemingly zero, are points where quantum fluctuations suppress the instability, i.e., no transition takes place. For x=1 saturation of T$_C$ is achieved and no further enhancement expected. In agreement with the experiment T$_C$ depends nonlinearly on x and does not follow a $1/\sqrt{M_1}$ dependence. The related soft mode is shown in figure 28 as a function of x and temperature.

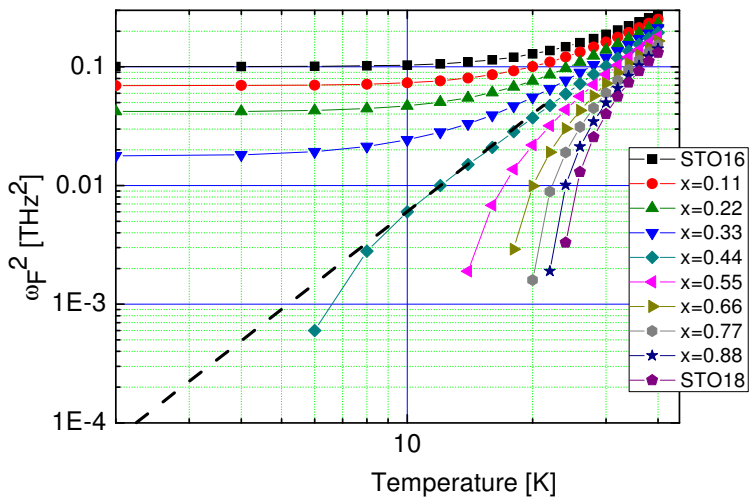

**Figure 28** Double logarithmic plot of the squared soft mode frequency as a function of temperature and for different x-values as indicated in the figure.

First it is noted that the x-dependent modes are not shifted in parallel. Second, for all values of x $\omega_F^2$ is nonlinearly dependent on temperature. Third, a distinctly different temperature dependence is observed for those x-values where a real instability takes place as compared to the quantum paraelectric systems. For values x>0.35 a linear $T^2$ dependence is obeyed which indicates the dimensionality crossover [36, 163]. Logarithmic corrections set in upon approaching $T_C$. The onset of the linear $T^2$ dependence shifts to higher values of T with increasing x, i.e., increasing $T_C$. This implies that the dimensionality crossover smoothly vanishes when $T_C$ increases. For those compounds where quantum fluctuations suppress the phase transition, a dimensionality crossover regime is again observed for x<0.35, where $\omega_F^2 \propto T^2$ over a broad temperature regime. This regime diminishes with decreasing x and is almost lost for x=0 where the onset temperature is beyond the temperature scale of figure 26. However, if increased temperatures are considered, the $\omega_F^2 \propto T^2$ regime is recovered and the crossover from d=4 to d=3 can clearly be observed. The black dashed line in figure 26 indicates the border between ferroelectric and quantum regime. The transition between both is not continuous but of first order since the limit $T_C$=0 is never reached, which rules out a quantum critical point.

The above results do not provide an answer to the controversy regarding the dynamics of the phase transition but only refer to the long wave length limit and do not address any local effects. These were investigated in detail by examining the acoustic mode dispersion and its evolution with temperature. With decreasing soft mode frequency, the acoustic and optic modes begin coupling at finite momentum q, whereby the acoustic mode develops an anomaly in its dispersion reminiscent of a ferroelastic instability. In order to highlight this anomaly, the acoustic mode dispersion is normalized to its value at T=200K where it still follows a harmonic $q^2$ dependence in the long wave length limit. The results of this procedure are shown in figure 29, where the temperature and momentum dependence of the normalized transverse acoustic mode are shown. While for T=180 and 160K the mode still behaves approximately harmonically, an anomaly develops at T=150K which becomes more pronounced with decreasing temperature. Simultaneously, the critical q value at which it appears shifts to the long wave length limit when approaching $T_C$. However, it never freezes and the q=0 limit is never reached, as should happen when the system forms a homogeneous ground state. These results provide fundamental evidence that dynamical clusters form around 150K with a length scale of approximately 6 lattice constants. With decreasing temperature the clusters increase in size and their dynamics slow down substantially. Around $T_C$ they reach a size of nearly 100 lattice constants and become quasi-static. Below $T_C$ the results are even more striking since the clusters shrink in size and their dynamics become faster again, even though these are still slow as compared to typical phonon frequencies. This behavior is distinctly different from SrTi$^{16}$O$_3$ where the cluster size continuously increases with decreasing temperature but always remains substantially smaller than in the present case. However, with increasing mass $M_1$ the cluster size increases as compared to the present case, and a homogeneous ferroelectric state is formed below $T_C$.

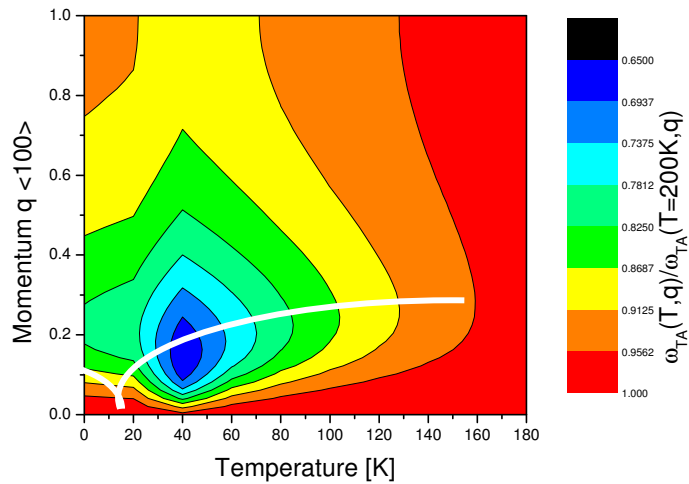

**Figure 29** Momentum and temperature dependence of the normalized transverse acoustic mode. The color code refers to its relative softening. The white lines refer to the wave vector $q_c$ where the acoustic mode exhibits an anomaly.

This observation evidences very clearly that the ferroelectric state of $SrTi^{18}O_3$ is not conventional. Rather, it is governed by a coexistence of dynamically distorted clusters and ferroelectrically polarized domains so that the ferroelectric state remains incomplete. This conclusion is supported by results from birefringence measurements where the coexistence of different symmetries in the low temperature state has been observed [164, 165]. In addition, NMR experiments in the high and low temperature phases [158, 166] can be understood within this approach since these also report a local symmetry breaking in both phases. Similarly Raman measurements [167, 168] and elastic and anelastic properties [169] provide evidence that the ferroelectric state of $SrTi^{18}O_3$ is unconventional.

## 7. Crossover between displacive and order / disorder dynamics

The mechanism of the phase transition in the perovskite ferroelectrics has been an important issue early on and has mostly been classified as either being of order / disorder or displacive type [117, 170 – 174]. Experimentally contradictory results have been obtained since long wave length testing tools like Raman scattering, infrared spectroscopy, dielectric measurements support the displacive case, whereas local probes like EXAFS, EPR, NMR demonstrate an order / disorder mechanism [54 – 61]. The observation of a soft mode in $BaTiO_3$, $SrTiO_3$, $KTaO_3$, $PbTiO_3$, $KNbO_3$ directly evidences the displacive character of the phase transition [16]. On the other hand strong diffuse X-ray scattering and Raman activity above the cubic to tetragonal phase transition of $BaTiO_3$ together with more detailed infrared and dielectric measurements and EPR spectroscopy [50, 175] reveal clearly order / disorder dynamics. From local probe measurements it was concluded that polarized clusters form above $T_C$ [59] and grow with decreasing temperature to coalesce into a ferroelectric state at $T_C$. An additional evidence for preformed polar domains has been obtained from the recent observation of an anomalous birefringence above $T_C$ [60, 61]. In a Brillouin scattering study [175] polar precursors have been detected in the paraelectric phase of $BaTiO_3$ which correlate with the softening of a longitudinal acoustic mode and appear already 80K above $T_C$. In a similar previous study acoustic phonon velocities have been measured [176] as a function of pressure and the anomalies seen there been interpreted as arising from anisotropic fields. Fluctuating polar clusters have also been detected by picosecond X-ray laser speckle technique [177] in the paraelectric phase. Not only $BaTiO_3$ exhibits these locally distorted regions well above $T_C$, there are similar observations for $PbTiO_3$ [178], $PbHfO_3$ [179, 180], $CdTiO_3$ [181], superlattices of $SrTiO_3/DyScO_3$ [182] and $SrTiO_3$ [153, 154] – all of them showing classical soft mode behavior. These data support the conclusion that polar nano – or microregions are common to perovskite ferroelectrics and are even generic and intrinsic implying that displacive and order / disorder dynamics always coexist but obey different time and length scales [31, 32,183].

Instead of using the SPA or the nonlinear solutions of the polarizability model, a molecular dynamics (MD) technique is applied with the model being extended to a 2D version (figure 30).

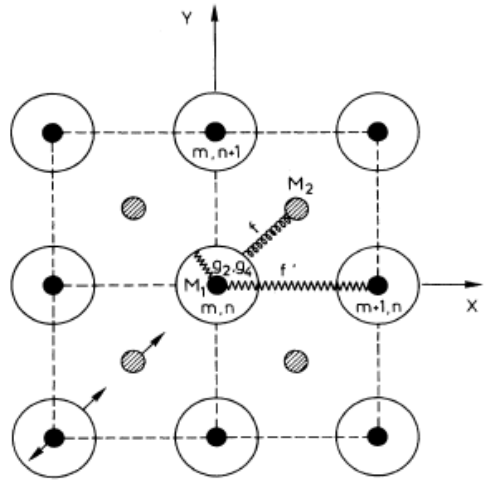

**Figure 30** 2D version of the polarizability model. The arrows on shells and cores in the left corner indicate the displacements in the ferroelectric phase [after Ref. 32].

As before $M_1, M_2$ refer to the polarizable cluster and rigid ion masses. The equations of motion are now more complex and the SPA avoided since MD techniques treat anharmonicities without approximations [32]. A constant temperature MD procedure has been employed to investigate the lattice dynamics in the vicinity of the phase transition. To control the temperature and maintain simultaneously stability, a predictor – corrector algorithm has been used [184] which includes the velocities in the predictor to calculate the positions such that the system's temperature can be adjusted by rescaling the particle velocities after each time step. The simulation has been performed on a square lattice of 32 x 32 unit cells with periodic boundary conditions. The runs have been carried through foe $2^{14}$ steps after zero time with a step size $\Delta \tilde{t} = 0.1 [|g_2|/M_2]^{1/2} t$ , and the temperature being given by: $\tilde{T} = (g_4 / g_2) k_B T$ . A vectorial order parameter has been

introduced and its nonzero value taken as evidence for a ferroelectric phase. The dynamical structure factor $S(q = 0, \omega)$ (i.e., the space time Fourier transform of the core-core displacement) at various temperatures above $\widetilde{T}_C$ is shown in figures 31.

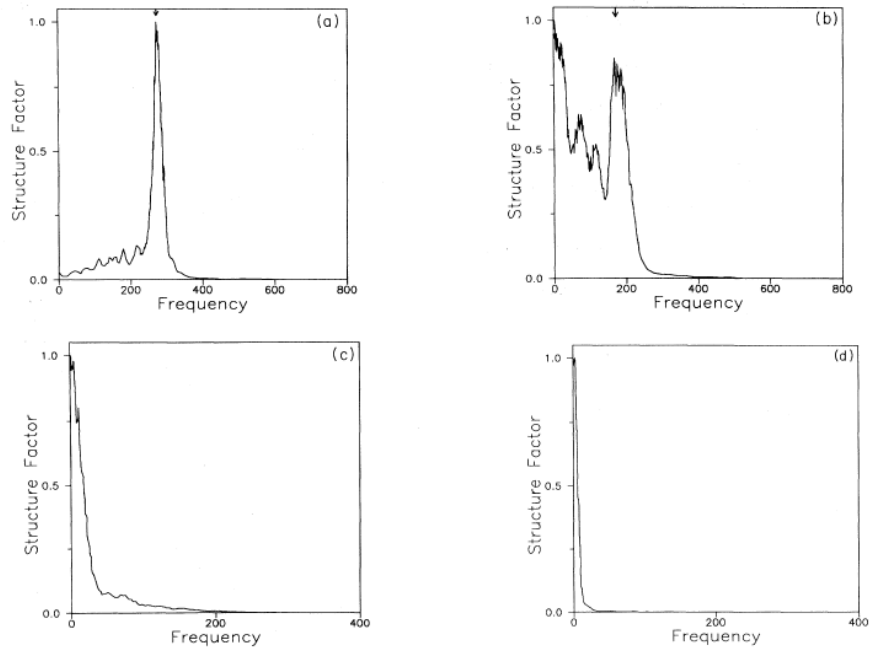

**Figure 31** Dynamical structure factor in the paraelectric phase for $\widetilde{T}/\widetilde{T}_C = 11.4$ (a), 2.9 (b), 1.4 (c), and 1.04 (d). The arrows indicate the SPA values of the soft mode [after Ref. 32].

With decreasing temperature the ferroelectric phonon peak softens and a central peak emerges gaining intensity with decreasing temperature. Its origin remains rather speculative [185], however, it can tentatively be attributed to precursor dynamics since from the MD simulation of a one component displacement model it could be assigned to travelling cluster waves [186]. The possible formation of precursor clusters has been investigated from the analysis of the evolution of the four ferroelectric domains being generated by the direction of the local polarization. The existence of correlated fluctuations in the cubic phase implies that the local order parameter has one of the four possible orientations which would lead to small locally ordered phases. Snapshots of instantaneous configurations are shown in figures 32. Fully uncorrelated polarizations exist (figure 32 (a)) for $\widetilde{T}/\widetilde{T}_C = 11.4$, while well below $\widetilde{T}_C$ (figure 32 (d)) the whole system has the same polarization except for isolated cell fluctuations. Between these two limits (figures 32 (b) and (c)) and at $\widetilde{T} > \widetilde{T}_C$ clusters of coherent polarization are present as expected from precursor dynamics.

Similar conclusions are reached from the analysis of the polarizability model based on the SPA [31]. The focus in that analysis concentrates on the soft mode temperature dependence and its coupling with the related acoustic mode analogous to the case of $SrTiO_3$. Since $T_C$ can be adjusted by varying $g_2$, a regime for $T_C$ has been selected lying outside the quantum critical and / or the saturation region. In the selected mean-field regime the soft mode is expected to follow the dependence $\omega_F^2 \approx (T - T_C)$. From the calculations it is seen that this behavior is not obeyed (figure 33) and deviations appear in the vicinity of $T_C$ and at high temperatures.

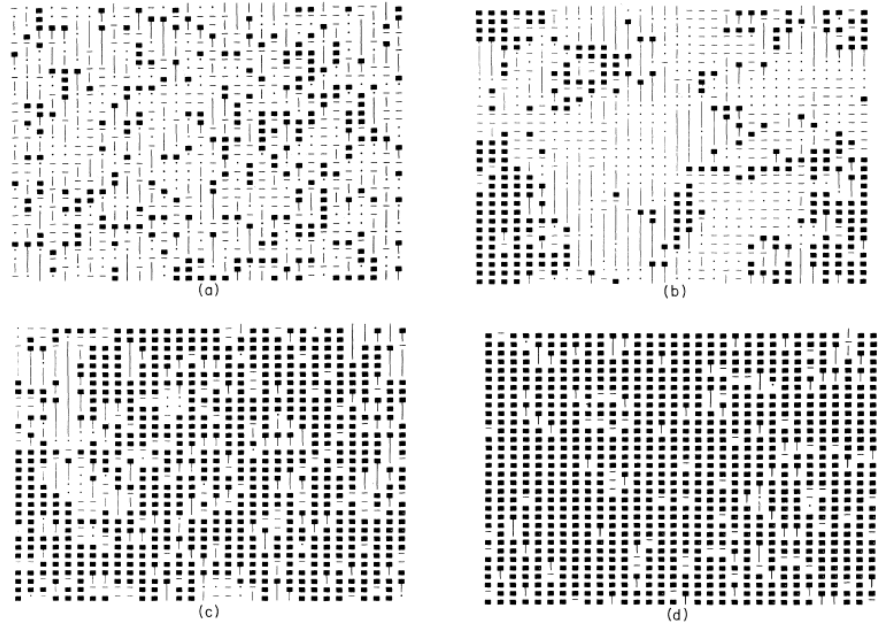

**Figure 32** Instantaneous cell polarization patterns in the paraelectric (a and b) and the ferroelectric phase (c and d) for $\tilde{T}/\tilde{T}_C = 11.4$ (a), 1.15 (b), 0.94 (c), and 0.66 (d). The different symbols refer to the four polarization directions [after Ref. 32].

For all three investigated values of $T_C$ the Curie constant increases with increasing $T_C$.

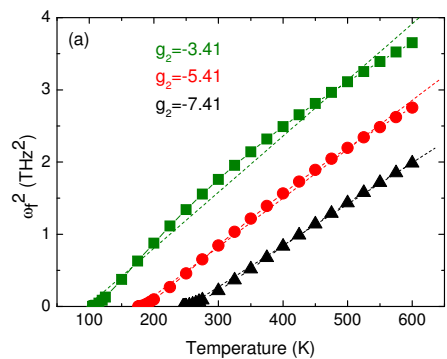

**Figure 33** The squared soft mode frequency $\omega_f^2$ as a function of temperature for three different values of $g_2$ as indicated in the figure.

The main purpose of calculating the soft mode frequency as a function of temperature is, however, to show that a long wave length instability exists as observed by many experiments. This finding primarily suggests that the transition described here is of the pure displacive type. That this conclusion is incomplete can be derived from the dispersion of the optic and acoustic modes, which start to couple at finite momentum.

As long as the coupling constant $g_T$ is large as compared to the nearest neighbor coupling constant, i.e., $T \gg T_c$, this coupling is temperature independent. If the opposite limit applies when T approaches $T_c$ but still being appreciably far away, a temperature and momentum dependent coupling sets in which gets more pronounced with decreasing temperature and induces momentum dependent anomalies in the acoustic mode dispersion. The critical momentum $q_c$ where these anomalies appear, provides information about the length scales and their local extensions in real space. In order to compare the three cases of different $T_c$'s studied here, the high temperature limit T=600K is used for the acoustic mode to compare the temperature evolution of the anomaly. This is displayed in figure 34 (a) where $q_c$ is shown as a function of normalized temperature.

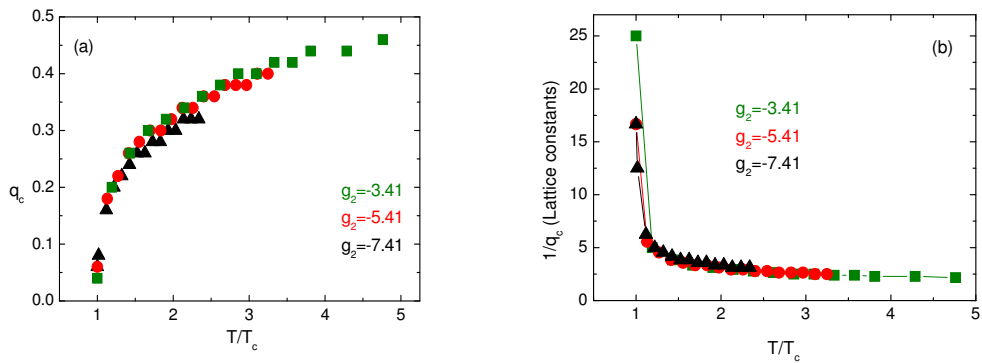

**Figure 34** (a) Dependence of the critical momentum $q_c$ on T/T$_C$ for different values of $g_2$ as indicated in the figure. (b) The critical length scale as a function of T/T$_C$ for different values of $g_2$ as indicated in the figure.

Obviously, for all values of T$_C$ $q_c$ lies on a universal curve. The farther T is above T$_C$ the larger is the momentum at which the anomaly is present. In addition, $q_c$ converges to a constant value, i.e., $q_c = 0.5$ for T$\gg$T$_C$, which signals the inherent tendency of the system to a doubling of the unit cell corresponding to a zone boundary instability. Indeed such an instability is observed in many perovskite ferroelectrics. Upon decreasing the temperature and approaching T$_C$ $q_c$ shifts to the long wave length limit indicating a coalescence of the local dynamics into a coherently polarized state. The related length scale of these dynamically distorted regions is shown in figure 34 (b) which emphasizes the divergent behavior appearing at T$_c$. It is interesting to note here, that a universal behavior is observed in all three cases even though the soft mode temperature evolution differs substantially from universality. Far above T$_C$ these regions are small and of the order of a few lattice constants only. When the temperature reaches about twice the transition temperature, their spatial dimensions start to grow to be almost doubled at T/T$_C$=1.5. When approaching T$_c$ further their growth is rapid and reaches 10 to 15 lattice constants at T/T$_C$=1.1 which corresponds in the investigated cases to a temperature window of 10 to 30K. A very different response appears when instead of changing the potential height the core-core coupling $f'$ is varied. In that case an extremely complex phase diagram is the consequence [183]. Yet, the common aspect of both approaches is that pretransitional local dynamics occur far above T$_C$. While the momentum anomaly reveals information about the length scale involved in the local dynamics as quantified above, the frequency value at the anomaly provides information about the time scale: For the long wave length soft mode $\omega_F$ the time scale is in the THz regime, for the precursor dynamics substantially slower time scales apply which are in the MHz to GHz region. This is shown in figure 35 where again these data are compared to the 600K dispersion of the acoustic mode.

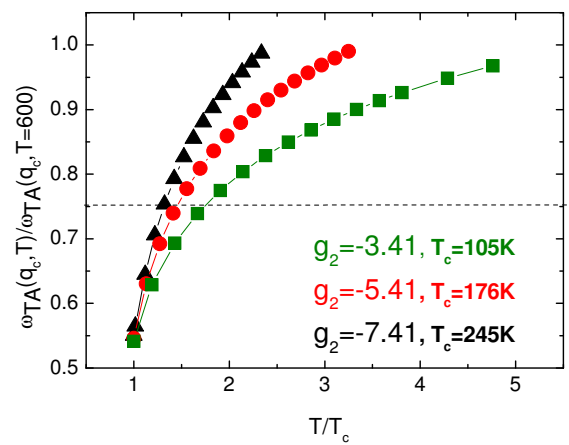

**Figure 35** Normalized acoustic mode frequency $\omega_{TA}(q_c,T)/\omega_{TA}(q_c,T=600K)$ versus T/T$_C$ for three different values of $g_2$.

Comparing the temperature dependence of the acoustic mode frequency at $q_c$ for the three different cases, universality is absent here. For the highest T$_C$ (245K) the frequency rapidly approaches the value of the harmonic mode whereas this is still substantially smaller in the case of T$_C$=105K even if the temperature is five times larger than T$_c$. In spite of these different temperature dependencies a common aspect of all three cases can be deduced from the result. By defining a minimum softening of 75% as compared to the 600 K mode frequency to be relevant to detect this softening experimentally (which corresponds to a domain size of ~ 5 lattice constants), the onset temperatures can be derived from figure 35. In all cases the precursor dynamics start to develop approximately 75 K prior to the long wave length instability. Even if a larger percentage is taken as the relevant one and the temperature scale for the local dynamics increases, a similar trend is observed. This means that the local dynamics develop always almost at the same temperature scale above T$_C$ in the investigated systems in full agreement with experimental data [60, 61, 187 − 189]. Another affirmation of the precursor scenario developed above comes from linear and nonlinear elastic and dielectric coefficients which have been measured for $BaTiO_3$ almost 30 years ago [190]. The analysis of the data has been done in terms of a thermodynamic functional [191] from which the temperature dependencies of the coefficients was derived. The comparison with the experiment showed that strong deviations between both exist which can only be explained if local symmetry breaking due to precursor dynamics takes place [31, 32].

Both of the above outlined analysis, the MD technique and the SPA of the polarizability model demonstrate that a displacive soft mode can coexist with precursor dynamics. This establishes the important conclusion that classification schemes in terms of order / disorder and displacive dynamics are misleading. It is also in support of different experimental results since length and time scales may differ considerably for both types and are thus only observable by specific tools.

## 8. Conclusions

The purpose of the review was to give a survey of the polarizability model, its essential ingredients, and its applications for perovskite ferroelectrics. This survey is not complete since a huge amount of work has also been done within the 3D model – not discussed here. This includes the lattice dynamics of $KTaO_3$ [39], structural instabilities of $KNbO_3$ and $KTaO_3$ [192 – 194] and superlattices of them [195], off-centering of Li ions in $Li_xK_{1-x}TaO_3$ [196], interface effects in ultrathin films [197], surface effects in $BaTiO_3$ [198] and atomistic modeling form first-principles [199]. Also, approaches based on first principles calculations and effective Hamiltonians are not included even though a large amount of work has been done in this area. Similarly, molecular dynamics studies and Monte Carlo methods have not been addressed. The reasons for these omissions are obvious from the title of this review, but also from the fact that it is impossible to review the broad range of literature in the field of ferroelectrics. However, it is worth mentioning that at present none of the above methods and models provides an *overall* description of the different types of ferroelectric perovskites (relaxors, order/disorder systems, displacive systems) consistently and in agreement with experiment. The above mentioned approaches do not rely on one single model as given by the polarizability model, but address each compound individually. Also, those approaches and methods are either restricted to the T=0K limit or fail to reproduce the correct transition temperatures and/or unit cell volumes. Another shortcome is their inability to capture the quantum limit, but always use is made of the phenomenological Barrett equation. The advantage of the polarizability model is not only its mostly analytically tractable form, but also the fact that essential parameters like the double-well potential defining quantities, are derived self-consistently. As such one must conclude that the core-shell coupling and its higher order terms are essentials which are absent in other approaches.

This coupling is a consequence of the specific role of nonlinearly polarizable ions (specifically the oxygen ion $O^{2-}$) and their function in structural phase transitions from paraelectric to polar and ferroelectric. The main results are related to the fact that earlier introduced classification schemes of ferroelectrics are not needed, since all specific features of the respective class emerge from a single model Hamiltonian. Crossovers between different dynamical regimes evolve from it, and coexistences have been shown to be present. An important issue relates to time and length scales which can differ strongly and are tested by different experiments. This can lead to seemingly controversial interpretations of data, which indeed are complementary. The nonlinearity appears as important ingredient also in the exact solutions of the model and gives rise to novel features related to domain walls, solitary and breather excitations.

The unconventional polarizability of the oxygen ion and its homologues is certainly not only playing an essential role for perovskite ferroelectrics but should also be taken into account when physics of other oxides are considered.

**Acknowledgement:** It is a great pleasure to acknowledge support, discussions with and active contributions from A. R. Bishop and H. Büttner. Besides of them a large community has contributed in obtaining the results presented in this review. These are J. Banys, H. Beige, G. Benedek, N. Dalal, A. Dobry, S. Kamba, M. Maglione, K. H. Michel, R. Migoni, K. A. Müller, K. Roleder, M. Sepliarsky, A. Simon, M. Stachiotti, K. Szot, and G. Völkel.

## 9. References

1. Wul B and Goldman J M 1945 *C. R. Acad. Sci. URRS* **46** 139; **49**, 177.
2. Wul B and Goldman J M 1946 *C. R. Acad. Sci. URRS* **51** 21.
3. Matthias B T 1949 *Phys. Rev.* **75** 177.
4. Matthias B T and Remeika J P 1956 *Phys. Rev.* **103** 262; 1949 *Phys. Rev.* **76** 1886.
5. Shirane G, Hoshino S and Suzuki K 1950 *Phys. Rev.* **80** 1105
6. Slater J C 1950 *Phys. Rev.* **78** 748.
7. Cochran W 1960 *Adv, Phys.* **10** 401.
8. Anderson P W 1960 *Fizika Dielektrikov* (Acad. Nauk, SSSR, Moscow).
9. Müller H 1940 *Phys. Rev.* **57** 829; *Phys. Rev.* **58** 565, 805.
10. Ginzburg V L 1945 *Zh. Eksp. Teor. Fiz.* **15**, 739; 1951 *Zh. Eksp. Teor. Fiz.* **19** 36.
11. Devonshire A F 1949 *Phil. Mag.* **40** 1040; 1951 *Phil. Mag.* **42** 1065.
12. Thomas H 1969 *IEEE Trans. Magn.* **5** 847; 1971 *Structural Phase Transitions and Soft Modes* (ed. Samuelsen E J, Universitetsvorleiget Oslo) p. 15.
13. Müller K A and Burkhard W 1979 *Phys. Rev. B* **19** 3593.
14. Barrett J H 1952 *Phys. Rev.* **86** 118.
15. Migoni R, Bilz H and Bäuerle D 1976 *Phys. Rev. Lett.* **37** 1155.
16. Lines M E and Glass A M 1977 *Principles and Applications of Ferroelectrics and Related Materials* (Clarendon, Oxford).
17. Bilz H, Benedek G and Bussmann-Holder A 1987 *Phys. Rev. B* **35** 4840.
18. Biltz H and Klemm W 1934 *Raumchemie der festen Stoffe* (Verlag Voss L, Leipzig).
19. Tessman G R, Kahn A H and Shockley W 1953 *Phys. Rev.* **92** 890.
20. Bilz H, Büttner H, Bussmann-Holder A, Vogl P 1987 *Ferroelectrics* **73** 493.
21. Bussmann A, Bilz H, Roenspiess R and Schwarz K 1980 *Ferroelectrics* **25** 343.
22. Watson R E 1958 *Phys. Rev.* **111** 1108.
23. Thorhallson G, Fisk C and Fraga S 1968 *Theor. Chem. Acta* **10** 388.
24. Prat R F 1972 *Phys. Rev. A* **6** 1735.

25. Cochran W 1960 *Adv. Phys.* **9** 387.
26. Cowley R A 1965 *Phil. Mag.* **11** 673.
27. Stirling W G 1972 *J. Phys. C* **5** 2711.
28. Bilz H, Bussmann A, Benedek G, Büttner H and Strauch D 1980 *Ferroelectrics* **25** 339.
29. Kugel G, Fontana M and Kress W 1987 *Phys. Rev. B* **39** 813.
30. Perry C H, Currat R, Buhay H, Migoni R M, Stirling W G and Axe J D 1989 *Phys. Rev. B* **39** 8666.
31. Bussmann-Holder A, Beige H and Völkel G 2009 *Phys. Rev. B* **79** 184111.
32. Stachiotti M, Dobry A, Migoni R M and Bussmann-Holder A 1993 *Phys. Rev. B* **47** 2473.
33. Bussmann-Holder A and Büttner H 1990 *Phys. Rev. B* **41** 9581.
34. Yamada Y and Shirane G 1969 *J. Phys. Soc. Jpn.* **36** 366.
35. Migoni R, Bilz H, Bussmann-Holder A and Kress W 1985 *Proc. Of the International Conference on Phonon Physics* (World Scientific, Singapore) p. 198.
36. Schmeltzer D 1983 *Phys. Rev. B* **28** 459.
37. Bussmann-Holder A and Bishop A R 2004 *Phys. Rev B* **70** 184303.
38. Bishop A R, Bussmann-Holder A, Kamba S and Maglione M 2010 *Phys. Rev. B* **81** 064106.
39. Macutkevic J, Banys J, Bussmann-Holder A and Bishop A R 2011 *Phys. Rev. B* **88** 184301.
40. Bussmann-Holder A, Bilz H and Benedek G 1989 *Phys. Rev. B* **39** 9214.
41. Comés R and Shirane G 1972 *Phys. Rev. B* **5** 1886.
42. Shirane G, Nethans R and Minkiewicz V G 1967 *Phys. Rev.* **157** 396.
43. Axe J D, Harada J and Shirane G 1970 *Phys. Rev. B* **1** 227.
44. Shirane G, Axe J D, Harada J and Remeika J P 1970 *Phys. Rev B* **2** 155.
45. Cowley R A 1964 *Phys Rev.* **134** A981.
46. Stirling W G and Currat R 1972 *J. Phys. C* **9** L519.
47. Stirling W G 1972 *J. Phys. C* **5** 2711.
48. Bussmann-Holder A, Köhler J, Kremer R K and Law J M 2011 *Phys. Rev. B* **83** 212102.
49. Bettis J L, Whangbo M-H, Köhler J, Bussmann-Holder A and Bishop A R 2011 *Phys. Rev. B* **84** 184114.
50. Müller K A and Berlinger W 1986 *Phys. Rev. B* **34** 6130.
51. Müller K A, Berlinger W, Blazey K W and Albers J 1987 *Solid State Comm.* **61** 21.
52. Müller K A and Fayet J C 1991 *Structural Phase Transitions II* (Springer Verlag, Berlin) p. 1.
53. Comes R, Lambert M and Guinier A 1968 *Solid State Comm.* **6** 715.
54. Zalar B, Laguta V V and Blinc R 2003 *Phys. Rev. Lett.* **90** 037601.
55. Zalar B, Lebar A, Seliger J, Blinc R, Laguta V V and Itoh M 2005 *Phys. Rev. B* **71** 064107.
56. Chaves A S, Barreto F C S, Nogueira R A and Zeks B 1976 *Phys. Rev. B* **13** 207.
57. Völkel G and Müller K A 2007 *Phys. Rev. B* **76** 094105.
58. Ravel B, Stern E A, Verdinskii R J and Kraizman V 1998 *Ferroelectrics* **206** 407.
59. Geneste G and Kiat J M 2008 *Phys. Rev. B* **77** 174101.

60. Ziebiňska A, Rytz D, Szot K, Górny G and Roleder K 2008 *J. Phys.: Cond. Mat.* **20** 142202.
61. Ko H-J, Kim T H, Roleder K, Rytz D and Kojima S 2011 *Phys. Rev. B* **84** 094123.
62. Benedek G, Bussmann-Holder A and Bilz H 1987 *Phys. Rev. B* **36** 630.
63. Calogero F and Degasperis A 1982 *Spectral Transform and Solitons: Tools to Solve and Investigate Nonlinear Evolution Equations* (North-Holland, Amsterdam, Volume I, II).
64. Calogero F 1983 *Statics and Dynamics of Nonlinear Systems* (Springer, Heidelberg) p. 7.
65. Rogers C and Shadwich W F 1982 *Bäcklund Transformations and their Applications* (Academic Press, New York).
66. Büttner H and Bilz H 1981 *Recent Developments in Condensed Matter Physics* (Plenum, New York, Vol. I).
67. Bilz H, Büttner H, Bussmann-Holder A, Kress W and Schröder U 1982 *Phys. Rev. Lett.* **48** 264.
68. Bishop A R, Krumhansl J R and Trullinger S E 1980 *Physica* (Utrecht) **1D** 1.
69. Pawley G S, Cochran W, Cowley R A and Dolling G 1966 *Phys. Rev. Lett.* **17** 753.
70. Cowley R A, Derby J K and Pawley G S 1969 *J. Phys. C* **2** 1916.
71. Hague M S and Hardy J R 1980 *Phys. Rev. B* **21** 245.
72. Shirane G, Axe J D, Harada J and Remeika J P 1971 *Phys. Rev. Lett.* **27** 1893.
73. Pintschovius L, Smith H G, Wababayashi N, Reichardt W, Weber W, Webb G W and Fisk Z 1983 *Phys. Rev. B* **28** 5866.
74. Allen P B 1972 *Solid State Comm.* **14** 937.
75. Bussmann-Holder A and Bishop A R 2004 *Phys. Rev. B* **70** 184303.
76. Aubry S 1997 *Physica D* **103** 201.
77. Sievers A J and Page J B 1994 *Dynamical Properties of Solids* (North-Holland, Amsterdam, Vol. VII) p. 157.
78. Flach S and Willis C R 1998 *Phys. Rep.* **295** 181.
79. Dolgov A S 1986 *Sov. Phys. Solid State* **28** 907.
80. Chen D, Aubry S and Tsironis G P 1996 *Phys. Rev. Lett.* **77** 4776.
81. Payrard M 1998 *Physica D* **119** 184.
82. Floria L M, Marin M, Martinez P J, Falo F and Aubry S 1996 *Europhys. Lett.* **36** 539.
83. Trías E, Mazo J J and Orlando T P 2000 *Phys. Rev. Lett.* **84** 741.
84. Binder P, Abraimov D, Ustinov A V, Flach S and Zolotaryuk Y 2000 *Phys. Rev. Lett.* **84** 745.
85. Sato M, Hubbard B E, Sievers A J, Ilic B, Czaplewski D A and Craighead H G 2003 *Phys. Rev. Lett.* **90** 044102.
86. Eisenberg H S, Silberberg Y, Morandotti R, Boyd H R and Aitchison J S 1988 *Phys. Rev. Lett.* **81** 3383.
87. Morandotti R, Peschel U, Aitchison J S, Eisenberg H S and Silberberg Y 1999 *Phys. Rev. Lett.* **83** 2726, 4756.
88. Mingeleev S V and Kivsheryu S 2001 *Phys. Rev. Lett.* **86** 5474.
89. Fleischer J W, Segev M, Efremidis M K and Christodoulides D M 2003 *Nature* (London) **422** 147.

90. Schwarz U T, English O and Sievers A J 1999 *Phys. Rev. Lett.* **83** 223.

91. Swanson B J, Brozik J A, Love S P, Strouse G F, Shreve A P, Bishop A R, Wang W Z and Salkola M I 1999 *Phys. Rev. Lett.* **82** 3288.

92. Kladko K, Malek J and Bishop A R 1999 *J. Phys.: Cond. Mat.* **11** L415.

93. Vulgarakis N K, Kalosakas G, Bishop A R and Tsironis G P 2001 *Phys. Rev. B* **64** 020301.

94. Kiselev S A, Lai R and Sievers A J 1998 *Phys. Rev B* **57** 3402.

95. Sievers A J and Takeno S 1988 *Phys. Rev. Lett.* **61** 970.

96. Page J B 1990 *Phys. Rev. B* **41** 7835.

97. Park S-E and Shrout T R 1997 *J. Appl. Phys.* **82** 1804.

98. Viehland D, Lang S, Cross L E and Wuttig M 1991 *Phil. Mag. B* **64** 335.

99. Cross L E 1987 *Ferroelectrics* **76** 214.

100. Burns G and Dacol F H 1983 *Phys. Rev. B* **82** 2527.

101. Smolenski G and Agranovskaya A 1960 *Sov. Phys. Solid State* **1** 1429.

102. Viehland D, Land S J, Cross L E and Wuttig M 1990 *J. Appl. Phys.* **68** 2916.

103. Randall C 1987 *PhD thesis, University of Essex.*

104. Akbas M K and Davies P K 1997 *J. American Ceram, Soc.* **80** 2933.

105. Burns G and Scott B A 1973 *Solid State Comm.* **13** 423.

106. Westphal V, Kleemann W and Glinchuk M 1992 *Phys. Rev. Lett.* **68** 847.

107. Blinc R, Doliňsek J, Gregorovič A, Zalar B, Filipič C, Kutnjak Z, Levstik A and Pric R 1999 *Phys. Rev. Lett.* **83** 424.

108. Blinc R, Laguta V V and Zalar B 2003 *Phys. Rev. Lett.* **91** 247601.

109. Bussmann-Holder A and Bishop A R 2004 *J. Phys.: Cond. Mat.* **16** L313.

110. Bussmann-Holder A, Bishop A R and Egami T 2005 *Europhys. Lett.* **71** 249.

111. Kleemann W, Dec J and Westwański B 1998 *Phys. Rev. B* **58** 8985.

112. Teslic S and Egami T 1998 *Acta Cryst. B* **54** 750.

113. Dmowski W, Vakhrushev S B; Jeong I-K, Hehlen M P, Trouw F and Egami T 2008 *Phys. Rev. Lett.* **100** 137602; Jeong I-K, Darling T W, Lee J K, Proffen Th, Heffner R H, Park J S, Hong K S, Dmowski W and Egami T 2005 *Phys. Rev. Lett.* **94** 147602.

114. Kamba S, Kempa M, Bovtun V, Petzelt J, Brinkmann K and Setter N 2005 *J. Phys.: Condens. Matter* **17** 3965

115. Vodopivec B, Filipič C, Levstik A, Holc J, Kutnjak Z and Beige H 2004 *Phys. Rev. B* **69** 224208

116. For a review see, e.g. Kamba S and Petzelt J 2004 *Piezoelectric Single Crystals and their Applications* (Penn. State University) p. 257.

117. Bruce A D, Müller K A and Berlinger W 1979 *Phys. Rev. Lett.* **42** 185.

118. Simon A, Ravez J and Maglione M 2004 *J Phys.: Cond. Mat.* **16** 963.

119. Farhi R, El Marssi M, Simon A, and Ravez 1999 *J Eur. Phys. J. B* **9** 599.

120. Ravez J, Broustera C, and Simon A 1999 *J. Mater. Chem,* **9** 1609.

121. Simon A, private comm.

122. Macutkevic J, Kamba S, Banys J, Brilingas A, Pashkin A, Petzelt J, Bormanis K, and Sternberg A 2006 *Phys. Rev. B* **74** 104106.

123. Kamba S, Nuzhnyy D, Bovtun V, Petzelt J, Wang Y L, Setter N, Levoska J, Tyunina M, Macutkevic J, and Banys J 2007 *J. Appl. Phys.* **102** 074106.

124. Hlinka J, Petzelt J, Kamba S, Noujni D, and Ostapchuk T 2006 *Phase Transitions* **79** 41

125. Mitsui T and Nomura S 1981 *Ferroelectrics and Related Substances,* Landolt Börnstein (Springer Verlag, Berlin, New York).

126. Gehring P M, Hiraka H, Stock C, Lee S-H, Chen W, Lee Z-G, Vakhrushev S B, and Chowdhuri Z 2009 *Phys. Rev. B* **79** 224109.

127. Gehring P M, Wakimoto S, Ye Z-G, and Shirane G 2001 *Phys. Rev. Lett.* **87** 277601.

128. Gehring P M, Park S-E, and Shirane G 2001 *Phys. Rev. B* **63** 224109.

129. Svitelskiy Oleksiy, La-Orauttapong Duangmanee, Toulouse Jean, Chen W, and Ye Z-G 2005 *Phys. Rev. B* **72** 172106.

130. Blinc R, Laguta V, and Zalar B 2003 *Phys. Rev. Lett.* **91** 247601.

131. Tsukuda S and Kojima S 2008 *Phys. Rev. B* **78** 144106.

132. Lushnikov S G, Fedoseev A I, Gvasaliya S N, and Kojima S 2008 *Phys. Rev. B* **77** 104122.

133. Laiho R, Lushnikov S G, Prokhorova S D, and Siny I G 1990 *Sov. Phys. Solid State* **32** 2024.

134. Siny I G, Lushnikov S G, Tu C-S, and Schmidt V H 1995 *Ferroelectrics* **170** 197.

135. Lushnikov S G, Ko J-H, and Kojima S 2004 *Appl. Phys. Lett.* **84** 4798.

136. Ahart M, Asthagari A, Ye Z-G, Dera P, Mao H-K, Cohen R E and Hemley R J 2007 *Phys. Rev. B* **75** 144410.

137. Smolenski G A, Yushin N K, and Smirnov S I 1985 *Sov. Phys. Solid State* **27** 801.

138. Kohutych A, Yevych R, Perechinskii S, Samulionis V,  Banys J, and Vysochanskii Yu 2010 *Phys. Rev. B* **82** 054101.

139. Bokov A A and Ye Z-G 2006 *J. Mater. Sci.* **41**, 31

140. For a recent review see, Samara G A 2003 *J. Phys. Cond. Mat.* **15** R367 and references therein.

141. Schäfer H, Sternin E, Stannarius R, Arndt M and Kremer F 1996 *Phys. Rev. Lett.* **76** 2177.

142. Tikhonov A N and Arsenin V Y 1977 *Solution of Ill-Posed Problems* (Wiley, New York).

143. Groetsch C W 1984 *TheSsolution of Tikhonov, Regularization for Fredholm Equation* (Pitman, London).

144. Provencher S W 1982 *Comput. Phys. Comm.* **27** 213.

145. Macutkevic J, Banys J and Matulis A 2004 *Nonlin. Anal. Model. Control* **9** 1.

146. Blinc R, 1960 *J. Phys. Chem. Solids* **13** 204 .

147. Blinc R and Zeks B 1972 *Adv. Phys.* **21** 693 .

148. Tokunaga M and Matsubara T 1966 *Prog. Theor. Phys.* **35** 857.

149. Bussmann-Holder A and Michel K 1998 *Phys. Rev. Lett.* **80** 2173.

150. N. Dalal, A. Klymachyov, and A. Bussmann-Holder 1998 *Phys. Rev. Lett.* **81** 5924.

151. Itoh M, Wang R, Inaguma Y, Yamaguchi T, Shan Y-J and Nakamura T 1999 *Phys. Rev. Lett.* **82** 3540.
152. Bussmann-Holder A, Büttner H and Bishop A R 2000 *J. Phys.: Cond. Mat.* **12** L115.
153. Bussmann-Holder A, Büttner H and Bishop A R 2007 *Phys. Rev. Lett.* **99** 167603.
154. Bussmann-Holder A and Bishop A R 2008 *Phys. Rev. B* **78** 104117.
155. Taniguchi H, Itoh M and Yagi T 2007 *Phys. Rev. Lett.* **99** 017602.
156. Takesada M, Itoh M and Yagi T 2006 *Phys. Rev. Lett.* **96** 227602.
157. Taniguchi H, Yagi T, Takesada M and Itoh M 2005 *Phys. Rev. B* **72** 064111.
158. Blinc R, Zalar B, Laguta V V and Itoh M 2005 *Phys. Rev. Lett.* **94** 147601.
159. Oppermann R and Thomas H 1975 *Z. Phys. B* **22** 387.
160. Schneider T, Beck H and Stoll E 1976 *Phys. Rev. B* **13** 1123.
161. Prosandeev S A, Kleemann W, Westwanski B and Dec J 1999 *Phys. Rev. B* **60** 14489.
162. Rubtsov A N and Janssen T 2001 *Phys. Rev. B* **63** 172101.
163. Rytz D, Höchli U T and Bilz H 1980 *Phys. Rev. B* **22** 359.
164. Dec J, Kleemann W and Itoh M 2005 *Phys. Rev. B* **71** 144113.
165. Zhang L, Kleemann W, Wang R and Itoh M 2002 *Appl. Phys. Lett.* **81** 3022.
166. Laguta V V, Blinc R, Itoh M, Seliger J and Zalar B 2005 *Phys. Rev. B* **72** 214117.
167. Shigenari Takeshi, Abe Kohji, Takemoto Tomohiko, Sanaka Osamu, Akaike Takashi, Sakai Yoshihide, Wang Ruiping and Itoh Mitsuru 2006 *Phys. Rev. B* **74** 174121.
168. Shigenari Takeshi, Abe Kohji 2008 *Ferroelectrics* **369** 117.
169. Scott J F, Bryson J, Carpenter M A, Herrero-Albillos J and Itoh M 2011 *Phys. Rev. Lett.* **106** 105502.
170. Scott J F 1974 *Rev. Mod. Phys.* **46** 83.
171. Lines M E 1972 *Phys. Rev. B* **5** 3690.
172. Eisenriegler E 1974 *Phys. Rev. B* **9** 1029.
173. Gillis N S and Koehler T R 1974 *Phys. Rev. B* **9** 3806.
174. Müller K A, Luspin Y, Servoin J L, and Gervais F 1982 *Journal de Physique Lettres* **43** L537.
175. Ko J-H, Kojima S, Koo T-Y, Jung J H, Won C J and Hur N J 2008 *Appl. Phys. Lett.* **93** 102905.
176. Ishidate T and Sasaki S 1989 *Phys. Rev. Lett.* **62** 67.
177. Tai R Z, Namikawa K, Sawada A, Kishimoto M, Tanaka M, Lu P, Nagashima K, Maruyama H, and Ando M 2004 *Phys. Rev. Lett.* **93** 087601.
178. Kwapuliński J, Pawelczyk M and Dec J 1997 *Ferroelectrics* **192** 307.
179. Kwapuliński J, Kusz J, Böhm H and Dec J 2005 *J. Phys.: Cond. Mat.* **17** 1825.
180. Kwapuliński J, Pawelczyk M and Dec J 1994 *J. Phys.: Cond. Mat.* **6** 4655.
181. Torgashev V I, Yuzyuk Y I, Shirokov V B, Lemanov V V, and Spektor I E 2005 *Phys. Solid State* **47** 337.

182.     Kužel P, Kadlec C, Kadlec F, Schubert J and Panaitov G 2008 *Appl. Phys. Lett.* **93** 052910.
183.     Bussmann-Holder A and Bishop A R 2009 *Ferroelectrics* **378** 42.
184.     Beeman D 1976 *J. Comp. Phys.* **20** 130.
185.     Schwabl F and Täubner U C 1991 *Phys. Rev. B* **43** 11112.
186.     Schneider T and Stoll E 1975 *Phys. Rev. B* **13** 1216.
187.     Kim Tae Hyun, Kojima Seiji, Park Kibog, Baek Kim Sung, Ko Jae-Hyeon 2010 *J. Phys.: Cond. Mat.* **22** 225904.
188.     Ohta Ryu, Zushi Junta, Ariizumi Takuma, Kojima Seiji 2011 *Appl. Phys. Lett.* **98** 092909.
189.     Kadlec Filip, Kadlec Christelle, Kužel Petr, and Petzelt Jan 2011 *Phys. Rev. B* **84** 205209.
190.     Beige H and Schmidt G 1982 *Ferroelectrics* **41** 173.
191.     Ljamov V E 1972 *J. Acoust. Soc. Am.* **52** 199.
192.     Sepliarsky M, Stachiotti M G and Migoni R L 1995 *Phys. Rev. B* **52** 4044.
193.     Sepliarsky M, Stachiotti M G and Migoni R L 1995 *Phys. Rev. B* **56** 566.
194.     Sepliarsky M, Stachiotti M G Migoni R L and Rodriguez C O 1999 *Ferroelectrics* **234** 9.
195.     Sepliarsky M, Phillpot S R, Wolf D, Stachiotti M G, Migoni R L 2001 *J. Appl. Phys.* **90**(9) 4509.
196.     Stachiotti M G and Migoni R L 1990 *J. Phys.: Cond. Mat.* **2** 4341.
197.     Sepliarsky M, Stachiotti M G and Migoni R L 2006 *Phys. Rev. Lett.* **96** 2006.
198.     Tinte S and Stachiotti MG 2001 *Phys. Rev. B* **64** 235403.
199.     Tinte S, Sepliarsky M, Stachiotti M G Migoni R L and Rodriguez C O 1999 *J. Phys.: Cond. Mat.* **11** 9679.